\documentclass[sigconf]{acmart}
\AtBeginDocument{%
  }

\copyrightyear{2025}
\acmYear{2025}
\setcopyright{cc}
\setcctype{by-sa}
\acmConference[CHI '25]{CHI Conference on Human Factors in Computing Systems}{April 26-May 1, 2025}{Yokohama, Japan}
\acmBooktitle{CHI Conference on Human Factors in Computing Systems (CHI '25), April 26-May 1, 2025, Yokohama, Japan}\acmDOI{10.1145/3706598.3714129}
\acmISBN{979-8-4007-1394-1/25/04}
\usepackage{algpseudocode}
\usepackage{algorithm}
\usepackage{soul}

\begin{document}

% \title{SplatOverflow: An Asynchronous Workflow for Maintaining Hardware} % <- Title A 
% \title{SplatOverflow: A Digital Representation for Hardware Workflows} % <- Title Bi 
% \title{SplatOverflow: A Tool for Talking About Hardware} % <- Title Bii
% \title{SplatOverflow: A Technique for Mediating Hardware Communication/Collaboration } % <- Title Bii
\title{SplatOverflow: Asynchronous Hardware Troubleshooting} % <- Title Biii


\author{Amritansh Kwatra}
\affiliation{%
  \institution{Cornell Tech}
  \city{New York}
  \country{USA}}
\email{ak2244@cornell.edu}

\author{Tobias Weinberg}
\affiliation{%
  \institution{Cornell Tech}
  \city{New York}
  \country{USA}}
\email{tmw88@cornell.edu}

\author{Ilan Mandel}
\affiliation{%
  \institution{Cornell Tech}
  \city{New York}
  \country{USA}}
\email{im334@cornell.edu}

\author{Ritik Batra}
\affiliation{%
  \institution{Cornell Tech}
  \city{New York}
  \country{USA}}
\email{rb887@cornell.edu}

\author{Peter He}
\affiliation{%
  \institution{Cornell University}
  \city{Ithaca}
  \country{USA}}
\email{ph475@cornell.edu}

\author{Fran\c{c}ois Guimbreti\`ere}
\affiliation{%
  \institution{Cornell University}
  \city{Ithaca}
  \country{USA}}
\email{fvg3@cornell.edu}

\author{Thijs Roumen}
\affiliation{%
  \institution{Cornell Tech}
  \city{New York}
  \country{USA}}
\email{tjr92@cornell.edu}

\renewcommand{\shortauthors}{Kwatra et al.}
\newcommand{\TODO}[1]{\textcolor{orange}{#1}}
\newcommand{\change}[1]{\textcolor{black}{#1}}
\newcommand{\urgentTODO}[1]{\textcolor{red}{#1}}

\begin{abstract}
As tools for designing and manufacturing hardware become more accessible, smaller producers can develop and distribute novel hardware. However, processes for supporting end-user hardware troubleshooting or routine maintenance aren't well defined. As a result, providing technical support for hardware remains ad-hoc and challenging to scale. Inspired by patterns that helped scale software troubleshooting, we propose a workflow for asynchronous hardware troubleshooting: SplatOverflow. 

SplatOverflow creates a novel boundary object, the SplatOverflow scene, that users reference to communicate about hardware. A scene comprises a 3D Gaussian Splat of the user's hardware registered onto the hardware’s CAD model. The splat captures the current state of the hardware, and the registered CAD model acts as a referential anchor for troubleshooting instructions. With SplatOverflow, remote maintainers can directly address issues and author instructions in the user’s workspace. Workflows containing multiple instructions can easily be shared between users and recontextualized in new environments. 

In this paper, we describe the design of SplatOverflow, the workflows it enables, and its utility to different kinds of users. We also validate that non-experts can use SplatOverflow to troubleshoot common problems with a 3D printer in a usability study. 

Project Page: \textcolor{blue}{\url{https://amritkwatra.com/research/splatoverflow}.}
\end{abstract}

\begin{CCSXML}
<ccs2012>
   <concept>
       <concept_id>10003120.10003121.10003129</concept_id>
       <concept_desc>Human-centered computing~Interactive systems and tools</concept_desc>
       <concept_significance>300</concept_significance>
       </concept>
 </ccs2012>
\end{CCSXML}

\ccsdesc[300]{Human-centered computing~Interactive systems and tools}

\keywords{Hardware Maintenance, Repair, Troubleshooting}

%%%%%%%% FIGURE 1 %%%%%%%%%%%%%%%%

%   \includegraphics[width=\textwidth]{figures/figone-submit.png}
%   \caption{(a)~ A SplatOverflow scene, comprising a 3D Gaussian Splat aligned and registered onto a digital CAD model, acting as a boundary object to facilitate asynchronous troubleshooting tasks for hardware. (b)~A remote maintainer using SplatOverflow to explore a local user's hardware and author instructions for them to follow. (c)~ A local user viewing the maintainer's instructions rendered as an overlay onto their hardware in their workspace and performing the specified action.}
%   \Description{}
%   \label{fig:figone}

\begin{teaserfigure}
  \includegraphics[width=\textwidth]{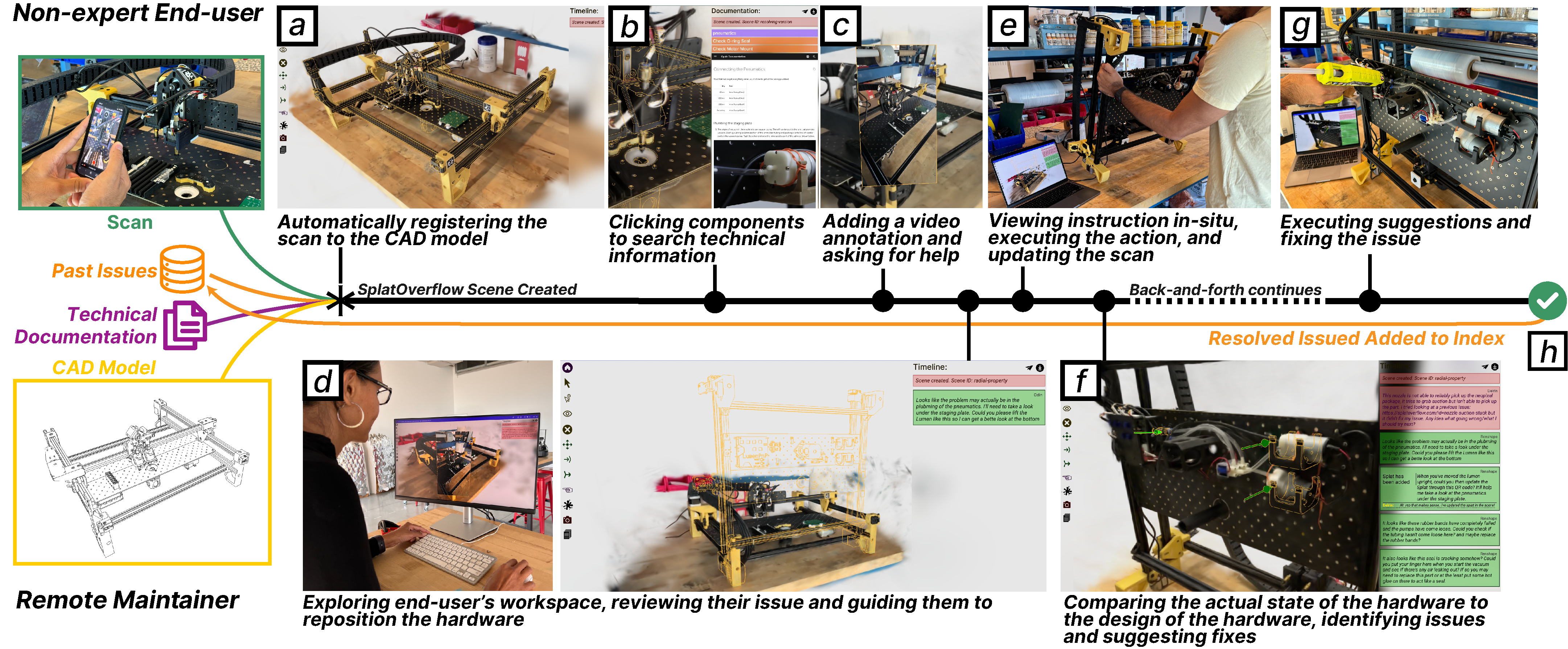}
  \caption{
    % \color{purple}
    A workflow illustrating a non-expert end-user using SplatOverflow to troubleshoot an issue with their hardware. (a) A SplatOverflow scene consisting of a user-captured scan of their hardware registered to the hardware's CAD model. (b) The user queries technical documentation and past issues associated with the hardware by clicking on components in the scene. (c) They request help from maintainers by sharing their SplatOverflow scene. (d) A maintainer explores the SplatOverflow scene and instructs the end-user to move their machine. (e) The local user sees instructions rendered as an overlay on their workspace and repositions their machine. (f) A maintainer compares the as-built hardware to the as-designed CAD model and suggests a solution. (g) The local user executes the suggestion and fixes the issue. (h) Once resolved, the issue, instructions, and deliberation are indexed back into a database of past issues for future users to reference.}
  \Description{
  Figure 1. This figure is a workflow diagram showing the different steps that go into troubleshooting with a SplatOverflow scene. It has eight components labeled (a) through (h) moving from (a) on the left to (h) on the right. At the left extreme of the figure there are four components that make up the SplatOverflow scene: A user-captured scan, existing technical documentation, a database of past issues and a CAD model of the hardware. Subfigure A depicts these elements coming together to create a SplatOverflow scene. Subfigure B shows how users can point and click on parts in the scene to retrieve their associated documentation or past issues that reference the part. Subfigure C shows how a user adds a video annotation and seeks out a maintainer's help. Subfigure D has two parts, on the left a maintainer reviews the end-user's issue on their screen, on the right, the SplatOverflow interface shows the suggestion authored by the maintainer. Subfigure E shows the user viewing the suggestion in situ, executing the fix, and updating the scene. Subfigure F shows the maintainer reviewing the updated scan, comparing the as-built hardware to the CAD design and proposing suggestions. The figure has a dashed line at this point, indicating that the issue could have more back-and-forth between the end-user and the maintainer. Subfigure G picks up when the end-user has identified the issue and executes the solution. Subfigure H loops back to the beginning, illustrating how resolved issues feed back into the database of past issues for future users to reference.
  }
  \label{fig:figone}
\end{teaserfigure}


\makeatletter
\def\@ACM@copyright@check@cc{}
\makeatother
\maketitle

\section{Introduction}

\begin{figure*}[h!]
    % \centering
    \includegraphics[width=\textwidth]{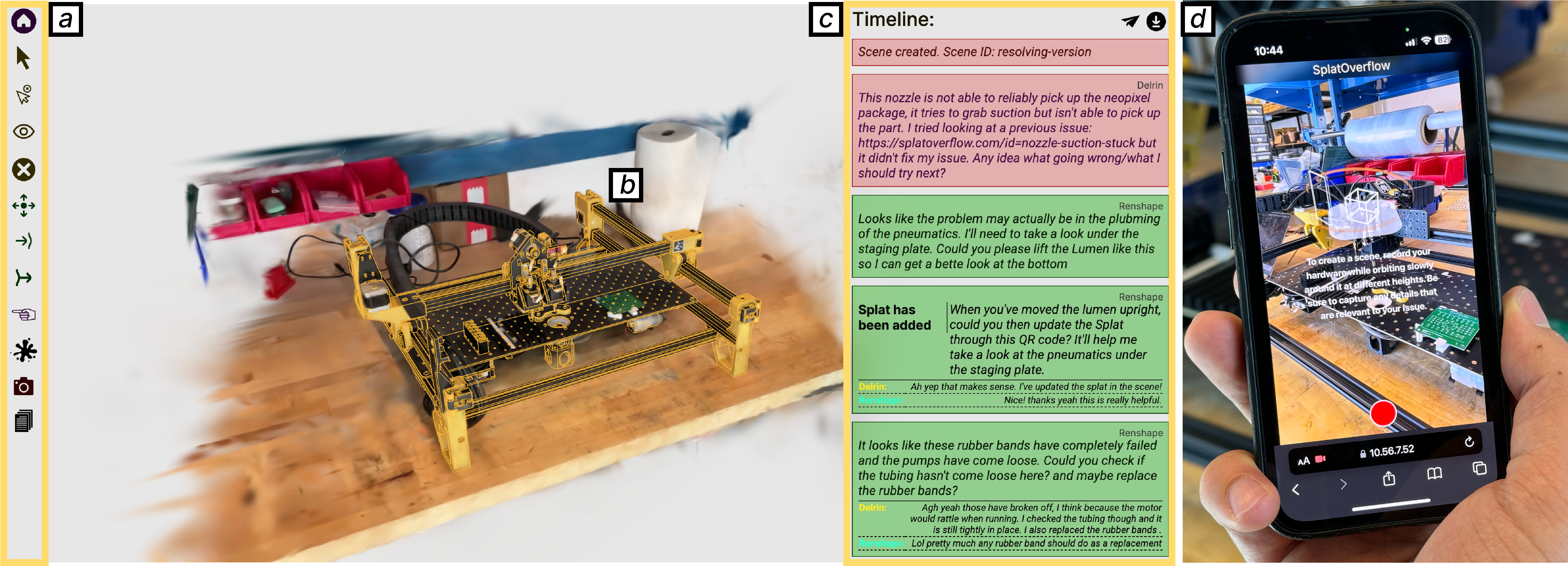}
    \caption{The components of SplatOverflow's web and mobile interface. (a) The palette of gestures SplatOverflow offers for selecting parts, guiding attention, and communicating actions. (b) The 3D Gaussian Splat of the hardware is aligned and registered onto the CAD model. (c) The timeline captures the troubleshooting interaction as a sequence of instructions and responses between users. (d) SplatOverflow's mobile interface allows local users to scan their hardware using a smartphone.}
    \Description{
    Figure 2. This figure has three subfigures in it: A, B, C, and D. The overall figure shows the SplatOverflow interface. The left box A highlights the menu of buttons that represent gestures SplatOverflow offers for selecting parts, guiding attention, and communicating actions. The middle label B shows the 3D Gaussian Splat of the hardware with the wireframe CAD model registered and aligned with the splat. The middle-right subfigure C shows SplatOverflow's timeline that captures the troubleshooting interaction as a sequence of instructions and responses between users. The farthest right subfigure D features a user holding a mobile phone with SplatOverflow’s mobile interface.
    }
    \label{fig:so-interfaces}
\end{figure*}

Hardware design tools have enabled small teams to develop and distribute novel machines and devices for niche applications. Examples of this type of hardware range from automation equipment for mid-scale manufacturing, such as the Lumen pick-and-place\footnote{\url{https://www.opulo.io/products/lumenpnp}}, to personal fabrication machines, like the Prusa MK3S\footnote{\url{https://help.prusa3d.com/tag/mk3s-2}}, to do-it-yourself gadgets like the Open Book\footnote{\url{https://www.oddlyspecificobjects.com/projects/openbook/}} e-reader. We refer to such products generally as \textit{hardware}. 
For hardware producers, turning a prototype into a viable product entails addressing challenges that stem from supporting a growing user base with varying technical expertise \cite{hodges_long_2019}. This paper focuses on one aspect of producing hardware: supporting end-user troubleshooting and maintenance. 

Detailed and thorough documentation is an essential part of developing a hardware product \cite{bonvoisin_what_2017, ackermann_toward_2008, hodges_long_2019}. However, documentation alone is insufficient to support the long tail of niche hardware issues consumers may encounter \cite{orr-talking-about-machines}. Instead, \citet{subbaraman_3d_2023} argue that maintenance should be considered a core part of owning this kind of hardware, and systems should be designed to support end-user troubleshooting and maintenance. We take inspiration from the infrastructure that supports software maintenance and troubleshooting workflows to examine how to create similar systems for hardware.

Platforms such as GitHub and StackOverflow have supported communities of software users by allowing them to help one another troubleshoot issues and by cataloging a history of past issues for anyone to reference. Crucially, these platforms rely on \emph{asynchronous} communication between users. This allows users to seek out help or provide suggestions without coordinating availability or scheduling meetings with one another. Eliminating this barrier has allowed distributed communities of users to flourish and made asynchronous modes the norm in software development \cite{yamauchi_collaboration_2000, ackerman_sharing_2013}. This contrasts the workflows HCI researchers have proposed for remote expert guidance and hardware troubleshooting, which are primarily designed for synchronous modes of communication \cite{feiner-virtualreplicas, gauglitz_integrating_2012, gauglitz_world-stabilized_2014}.

The utility of asynchronous workflows for troubleshooting is due, in part, to how they facilitate communication between users: via references to shared digital artifacts. These artifacts resemble boundary objects \cite{star_boundary_objects, ackerman_sharing_2013, lutters_beyond_2007} that serve as mechanisms for communicating context and ideas online. In StackOverflow, for example, issues are accompanied by segments of code that the user is writing. Suggestions are then made as references to lines of code the user shared. The boundary object (in this case, lines of code) captures the user's context and scaffolds the asynchronous communication with others. Importantly, this context is identical for those contributing to troubleshooting the issue and future users referencing the issue. As a result, a larger group of current and future users benefit from the knowledge gleaned in the interaction \cite{yamauchi_collaboration_2000}. 

Hardware needs a robust boundary object to support the asynchronous modes of communication necessary to scale maintenance and troubleshooting infrastructure. We present SplatOverflow, a system that enables asynchronous hardware troubleshooting through a novel boundary object: a SplatOverflow scene. A SplatOverflow scene captures the as-built hardware through a scan (a 3D Gaussian Splat \cite{kerbl_3d_2023}) and renders as-designed details by aligning the scan to a shared CAD model of the hardware. 

% The scan lets a remote user independently navigate a local user's environment, inspect issues with the as-built hardware, and manipulate assemblies to convey instructions. The CAD model lets local users reference parts, query technical documentation, and retrieve past issues without knowing hardware-specific terminology. Interactions with the SplatOverflow scene are captured in a timeline that captures the history of instructions and discussions relating to a hardware issue. 

The construction of a SplatOverflow scene presents distinct advantages to local users (who have the hardware in front of them), remote \change{maintainers}, and future users who may encounter similar hardware issues.
Local users benefit from access to technical documentation linked to the CAD model and the ability to describe issues on their hardware by pointing and clicking on parts rather than knowing hardware-specific vocabulary.
Remote \change{maintainers} benefit from seeing the local user's issue registered onto the familiar CAD model and being able to freely move through the scene to inspect the hardware in the local user's environment. 
Finally, future users benefit from being able to retrieve and replay solutions to past issues without seeking out support. Instead, SplatOverflow re-contextualizes instructions from a past issue by overlaying them onto the current user's environment.
This ability to accommodate multiple users and \change{maintainers} is essential to how SplatOverflow can help support scaling the maintenance effort for hardware. 

In this paper, we present the design of SplatOverflow, demonstrate how it can be used to troubleshoot issues on different kinds of hardware, and validate that non-expert users can use SplatOverflow to troubleshoot hardware issues.
\section{Walkthrough} \label{sec:walkthrough}

We demonstrate how a \change{non-expert} local \change{end-}user can troubleshoot a complex issue on the Lumen v3 Pick-and-Place machine using SplatOverflow. The Lumen is an automation machine designed to assemble PCBs by picking up surface-mount electrical components with a vacuum nozzle and placing them on circuit boards. In this scenario, the local user's machine can \change{not} reliably pick up parts with one of the two vacuum nozzles. They use SplatOverflow to capture the issue they are facing and receive suggestions from a remote maintainer. \change{Figure \ref{fig:figone} depicts a high-level visual illustration of this process}.

\subsection{Creating a SplatOverflow Scene}
The user opens SplatOverflow on their smartphone to capture the issue they are facing. Figure \ref{fig:so-interfaces}(d) shows SplatOverflow's capture interface. They record a roughly one-minute-long video, filming the Lumen from multiple angles. After recording, they upload the video, and SplatOverflow generates a scene. 

% \begin{figure}
%     \centering
%     \includegraphics[width=\linewidth]{figures/so-scene-2.png}
%     \caption{(a)~A 3D Gaussian Splat \protect\cite{kerbl_3d_2023} of the local user's Lumen v3. (b)~A SplatOverflow Scene, comprising a 3D Gaussian Splat aligned onto a CAD model. (c)~The CAD model is rendered as a yellow wireframe to preserve the rendering of physical details in the local user's hardware present in the splat.}
%     \label{fig:so-scene}
% \end{figure}

After two minutes, SplatOverflow completes generating a \textit{low-resolution} scene. This scene is generated using down-sampled video, which yields a lower quality splat. However, this lower-resolution scene allows the user to quickly begin using SplatOverflow. After roughly four minutes from the original upload, SplatOverflow finishes generating the \textit{full-resolution} scene containing the user's Lumen registered onto its CAD model. SplatOverflow removes the background regions of the splat by default to preserve the local user's privacy. What's left is a splat of the user's hardware and parts of the immediate work surface on which it is placed. Figure \ref{fig:so-interfaces}(a-c) shows the scene as it is rendered in the browser. 

\begin{figure}[b!]
    \centering
    \includegraphics[width=\linewidth]{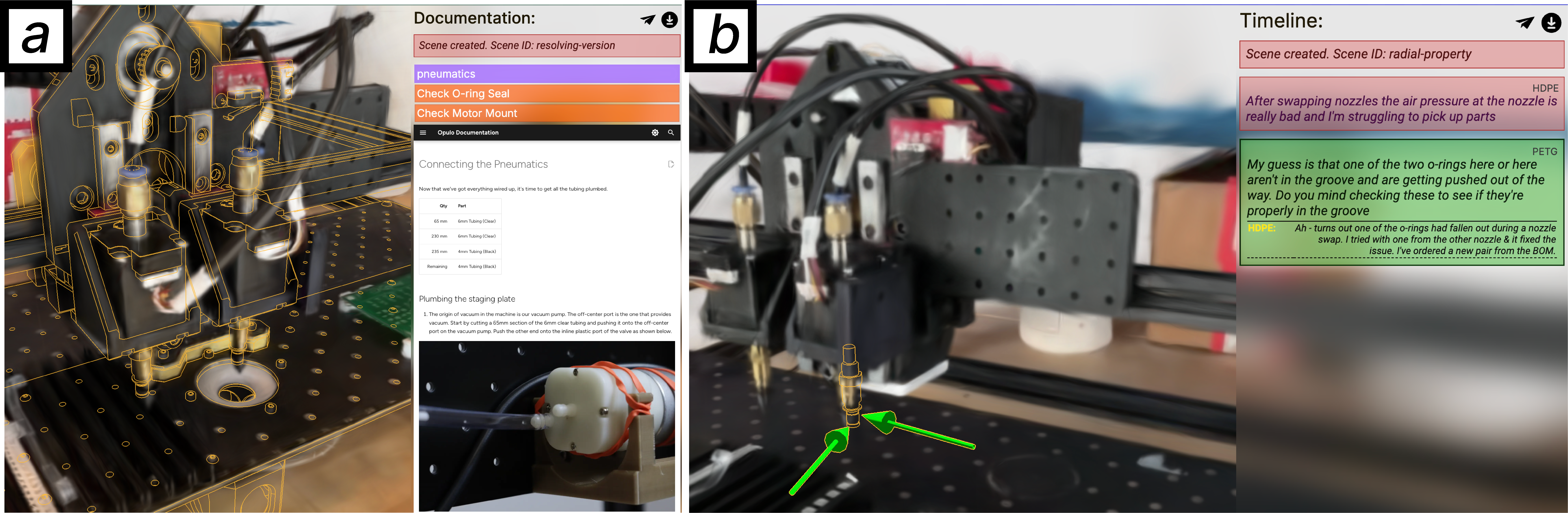}
    \caption{(a)~Using the CAD model to visually query technical documentation \change{and past issues}. When the user clicks on the nozzle, SplatOverflow retrieves relevant sections from the assembly documentation \change{and past issues referencing the part}. \change{Links to documentation} are rendered in purple and previewed in the timeline as an iframe. \change{Past issues are rendered in orange and can be recontextualized within the scene}. (b)~The local user \change{recontextualizes} a previous issue referencing the nozzle's poor suction; viewing the suggestions from a prior overlaid onto their machine.}
    \Description{
Figure 3. This figure has two subfigures in it: A and B. The left subfigure shows an example of the SplatOverflow documentation query interface. It features a user clicking on the nozzle of the machine in SplatOverflow and the corresponding nozzle documentation being displayed on the right side of the screen. The right subfigure shows an example of asynchronous collaboration using SplatOverflow’s timeline feature. With the user selecting the nozzle of the pick and place machine. Various past issues other users have had are displayed on the right side of the interface. The figure also features two arrow annotations pointing towards the nozzle being overlaid on the user’s machine, with the arrows being from a past user’s suggestions. 
    }
    \label{fig:so-docs-query}
\end{figure}

% In the previous issue, the maintainer indicated poor seating of two o-rings on the nozzle may be the issue. The local user tries this suggestion, but the o-rings are seated well, so they ask for more help.

\subsection{Visually Querying Documentation and Past Issues}
First, the local user examines any technical documentation for Lumen to find a solution. They select SplatOverflow's documentation tool and click on the nozzle in the scene. Figure \ref{fig:so-docs-query}(a) shows how SplatOverflow then displays sections of the technical documentation and past SplatOverflow issues referencing the nozzle assembly. The user scrolls through the retrieved documentation but does \change{not} find any suggestions about troubleshooting poor suction.

Next, the user sees a past issue that references a faulty nozzle. They open the issue, and SplatOverflow re-contextualizes the instructions from that scene onto their hardware. Figure \ref{fig:so-docs-query}\change{(b)} shows the solution to a \change{past} issue recontextualized in the local user's SplatOverflow scene. The user walks through the solution, which had to do with fixing the O-rings that seal the vacuum nozzle. They check that their O-rings are \change{not} dislodged and confirm this is not the source of their problem. They then decide to capture more details about the issue and ask for help. 

\begin{figure}[b!]
    \centering
    \includegraphics[width=\linewidth]{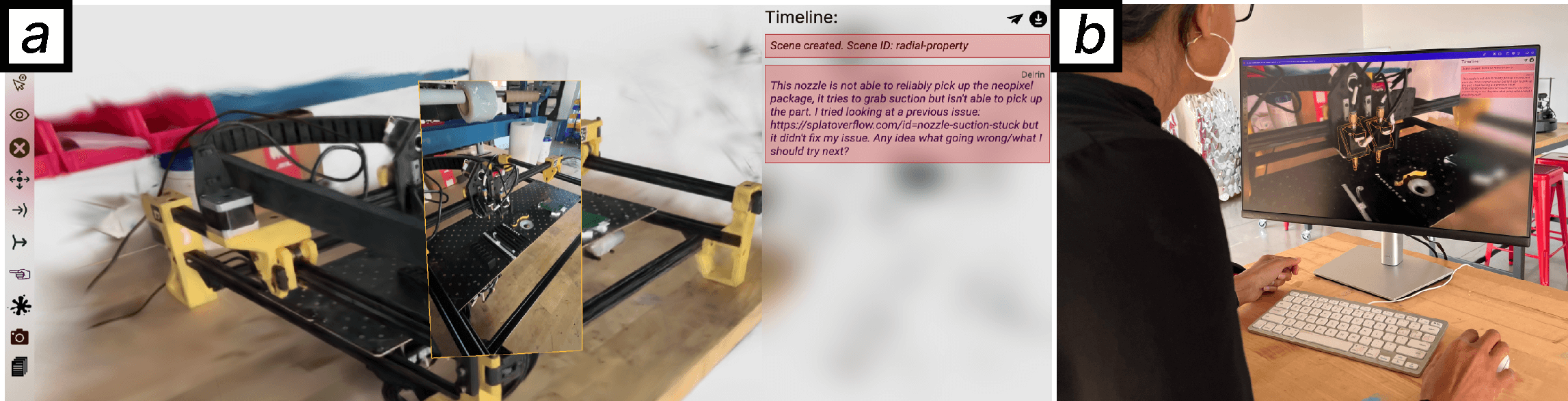}
    \caption{(a)~The local user's video is placed as a floating screen in the SplatOverflow scene and aligned to the same perspective it was filmed in. (b)~The remote maintainer reviews the local user's issue by inspecting the vacuum nozzle in the splat and CAD model.}
    \Description{
    Figure 4. This figure has two subfigures in it: A and B. The left subfigure A shows a local user’s video annotation rendered in the SplatOverflow scene of a pick and place machine. The video is placed as a floating screen in the 3D Scene and aligned to the same perspective it was filmed in. The right subfigure B shows the remote maintainer inspecting the vacuum nozzle in the SplatOverflow scene.
    }
    \label{fig:so-video-embed}
\end{figure}

\subsection{Asking for Help}
To share their issue with a remote maintainer, the local user adds a description of the problem they are facing. They also add a video \change{annotation} of the nozzle failing to pick up a component. SplatOverflow places the video into the scene to match the filming perspective, as shown in Figure \ref{fig:so-video-embed}\change{(a)}. The video describes the dynamic aspects of the issue that cannot be captured in the scan of the workspace. The local user then posts the issue and waits for a remote maintainer to view it and offer suggestions.

\subsection{Providing Guidance}
After a few hours, the remote maintainer reviews the SplatOverflow issue and offers feedback. Figure \ref{fig:so-video-embed}\change{(b)} shows the maintainer reviewing the issue and inspecting the nozzle assembly in the scene. The maintainer cannot see any problems with the nozzle assembly and suspects that the source of the issue may be the pneumatics under Lumen's staging plate. However, the underside of the staging plate is not visible in the machine's current orientation. 

\begin{figure}[t!]
    \centering
    \includegraphics[width=\linewidth]{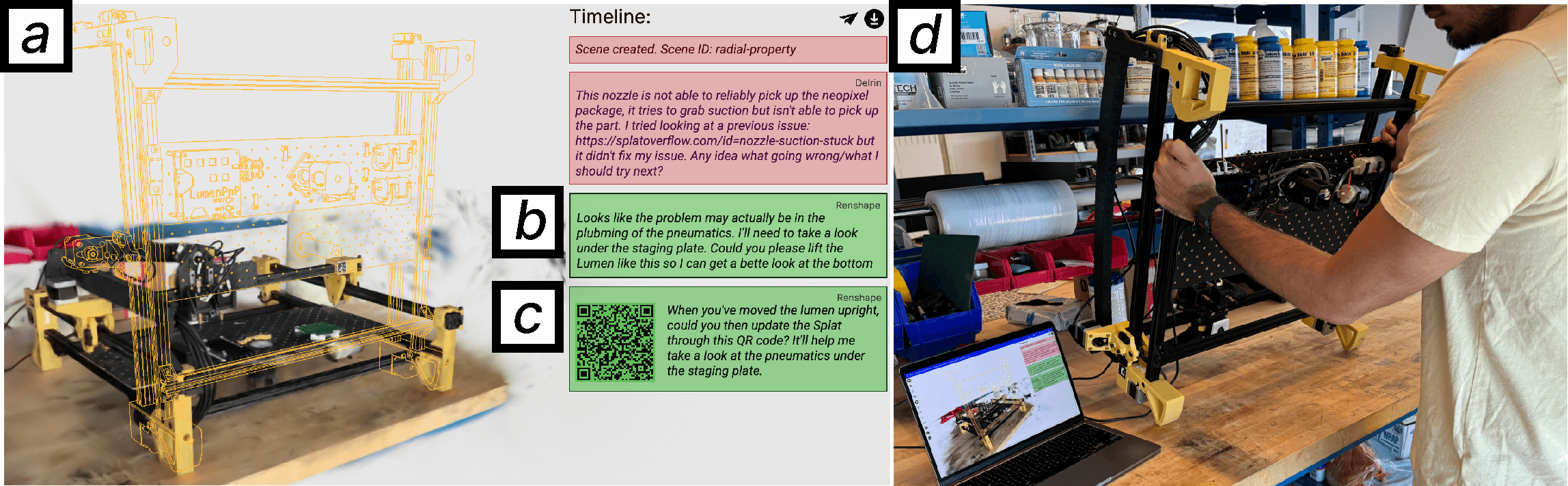}
    \caption{A reply from the remote maintainer asking the user to re-orient the machine to a new position and then update the scene with footage of the underside of the machine. \change{(a) T}he wireframe corresponding to the final position the machine should be in. \change{(b) T}he timeline element explaining why the movement needs to be made. When the local user clicks \change{this element}, the CAD model will animate to show how the machine should be moved. (c) The request for a new splat is rendered for the local user. The QR code links them to the mobile capture interface, where they can update the scene the QR code links to. (d) The local user performing the action specified by the maintainer and preparing to update the splat.}
    \Description{
    Figure 5. This figure has four subfigures in it: A, B, C, and D. The left subfigure A shows the wireframe corresponding to the final position a pick and place machine should be in for maintenance. The middle subfigure B shows a green button in the SplatOverflow timeline that describes the issue and also plays a CAD animation on how to move the machine once clicked. The other middle subfigure C is a green popup in the timeline that offers a QR code for the local user to capture a new splat of a machine in the new orientation described in B. The right subfigure D shows a local user performing the action specified by the maintainer and preparing to update the splat.
    }
    \label{fig:so-reorienting}
\end{figure}

The maintainer directs the local user to reorient the machine \change{by manipulating the CAD model in SplatOverflow} and requests that they update the captured splat after moving it. The move instruction is visualized by animating the CAD model to move from the original position to a new target position. The request to update the splat includes a QR code pointing the user to SplatOverflow's mobile capture interface for adding a splat to an existing scene. Figure \ref{fig:so-reorienting} shows the instructions authored by the maintainer in the timeline and how each is rendered in SplatOverflow. 

\begin{figure}[b!]
    \centering
    \includegraphics[width=\linewidth]{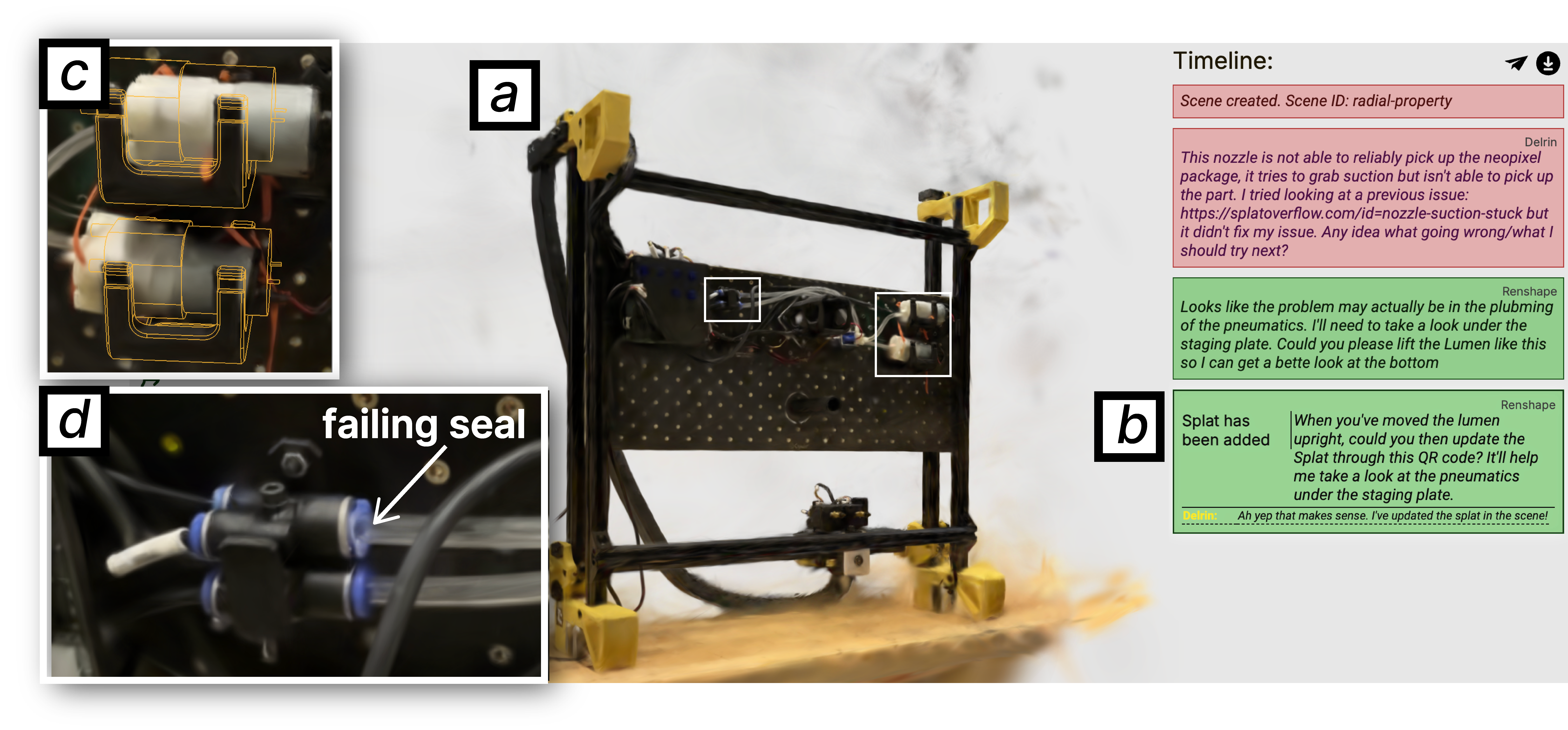}
    \caption{(a)~\change{T}he updated splat rendered in the browser. The machine is now standing upright as the maintainer requested, and the pneumatics in the staging plate are visible. (b)~\change{T}he timeline is updated to indicate that a new splat has been added, replacing the QR code with a success message. (c)~ A close-up of the vacuum pumps dislodged with their positions in the CAD model overlaid. (d)~ A close-up of the pneumatic seal that the maintainer is concerned about.}
    \Description{
    Figure 6. This figure has four subfigures in it: A, B, C, and D. The middle subfigure A shows an update of the previous splat with the pick and place machine now in the orientation specified by the remote maintainer. The right subfigure B shows that the SplatOverflow timeline has been updated to indicate that a new splat has been added. The upper left subfigure C shows a close up that highlights a part in the splat that the remote maintainer is concerned about, the vacuum pumps on the underside of the machine. The lower left subfigure D shows a close up that highlights aother part in the splat that the remote maintainer is concerned about a pneumatic fitting.
    }
    \label{fig:so-recapture-inspecting}
    % \description{} 
\end{figure}

\subsection{Understanding Suggestions}
The local user reviews the suggestions made by the remote maintainer. They see that the maintainer has asked them to move their machine. Clicking the instructions in the SplatOverflow timeline shows the user that they need to stand the machine on its back legs. Figure \ref{fig:so-reorienting}(d) shows the user following the maintainer's instructions by moving the machine and updating their splat. The user scans the QR code in the timeline event and captures a new splat showing the underside of the staging plate. Once the splat is updated, the scene contains two splats, with the machine in different orientations. Figure \ref{fig:so-recapture-inspecting}(a) shows the updated scene with the Lumen standing upright and the pneumatics easily visible for the remote maintainer to inspect. The splat captures the state of the wiring and routed tubes that are not included in the machine's CAD model.

\begin{figure}
    \centering
    \includegraphics[width=\linewidth]{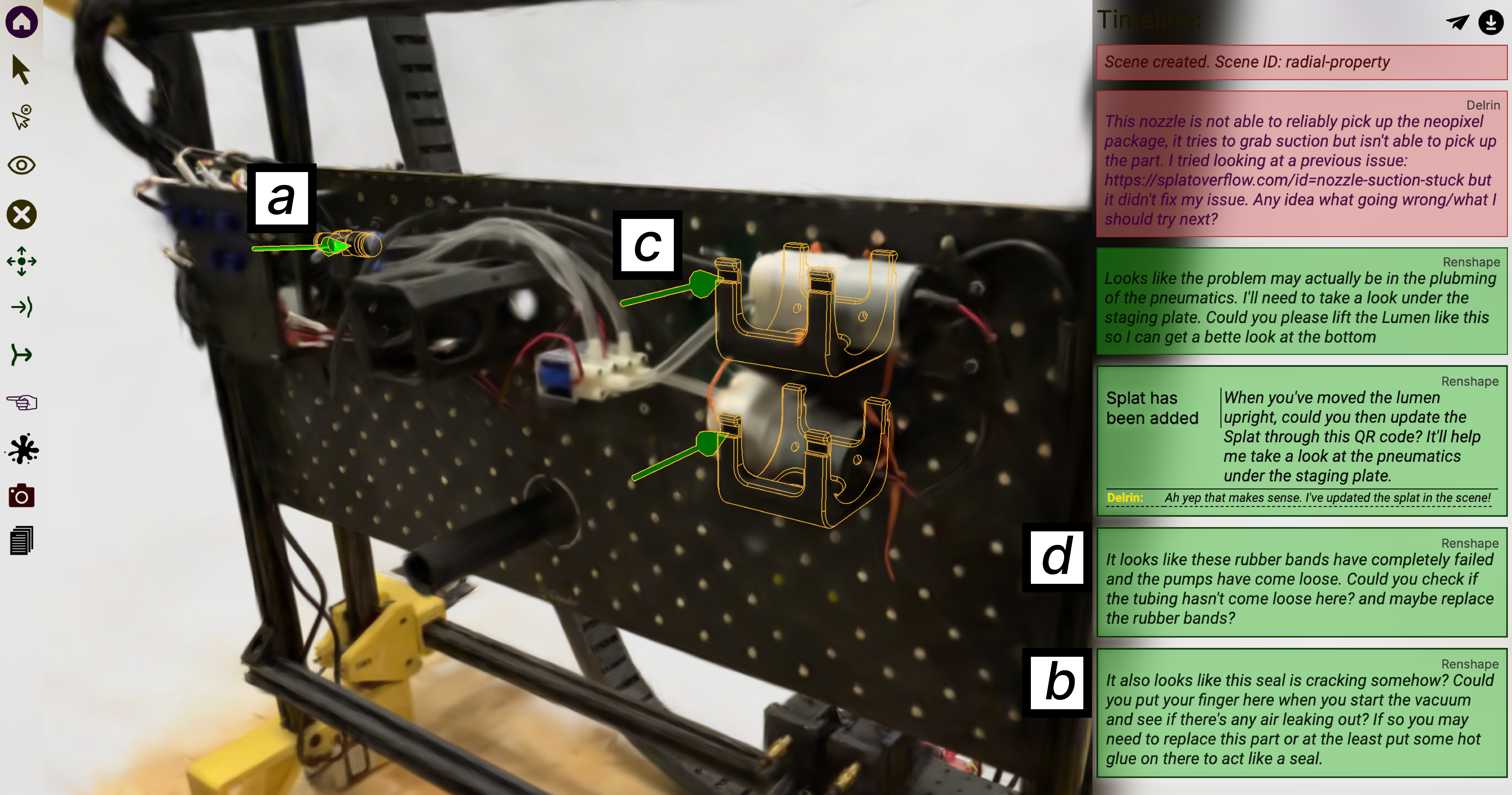}
    \caption{The remote maintainer makes two sets of suggestions. First, they indicate that based on the splat, one of the fittings appears to be failing. They instruct the user on how to test the fitting and offer a stop-gap solution if it has failed. (a) \change{T}he visualization indicates which part the remote maintainer references. (b) \change{T}he timeline element with the relevant instruction. Next, the remote maintainer notices that the pumps have come loose and asks the local user to check if the fitting is still secure. They also indicate that the local user should replace the deteriorated rubber bands. (c) \change{T}he visualization overlaid onto the pump mounts. (d) \change{T}he timeline element with the relevant instruction.}
    \Description{
    Figure 7. This figure has four subfigures in it: A, B, C, and D. Subfigure A is a green arrow pointing towards a part that the remote maintainer references. Subfigure B is a box in the timeline with the relevant instruction for the part highlighted in Subfigure A. Subfigure C shows the CAD wireframe overlaid onto missing parts in the Splat with Subfigure D giving relevant instructions on fixes.
    }
    \label{fig:so-making-suggestions}
\end{figure}

\subsection{Back and Forth Communication} 
The maintainer sees the updated splat and examines the pneumatics under the staging plate. Using the splat, they inspect the routing of tubes (not modeled in the CAD) and assess the condition of components that may be degrading. After orbiting around the model, the maintainer feels confident that the wiring is correct but notices two concerns. First, the rubber bands holding the pumps have wholly degraded, as shown in \change{Figure} ~\ref{fig:so-recapture-inspecting}\change{(c)}. Second, the blue fitting on one of the pneumatic fittings has come loose and will likely not provide an adequate seal, as shown in \change{Figure} ~\ref{fig:so-recapture-inspecting}\change{(d)}. Both are potential sources of the issue, so the remote maintainer instructs the user to fix both. They use SplatOverflow's \textit{pointing} gesture to attach comment\change{s} to the pneumatic fitting \change{and} the vacuum pumps. They instruct the local user to test if the tubes leaving the air pump are still securely attached and to replace the rubber bands when they have the chance. Next, the maintainer guides the user through testing if air escapes the fitting when the pump is turned on. The maintainer asks the user to place their finger on the seal and feel for air when the pump is on. In case the seal leaks air, the maintainer suggests that the local user apply some hot glue as a stop-gap solution while waiting for a replacement part. Figure \ref{fig:so-making-suggestions} shows how these \change{instructions are} rendered in the timeline and within the scene. 

\begin{figure*}
    \centering
    \includegraphics[width=\textwidth]{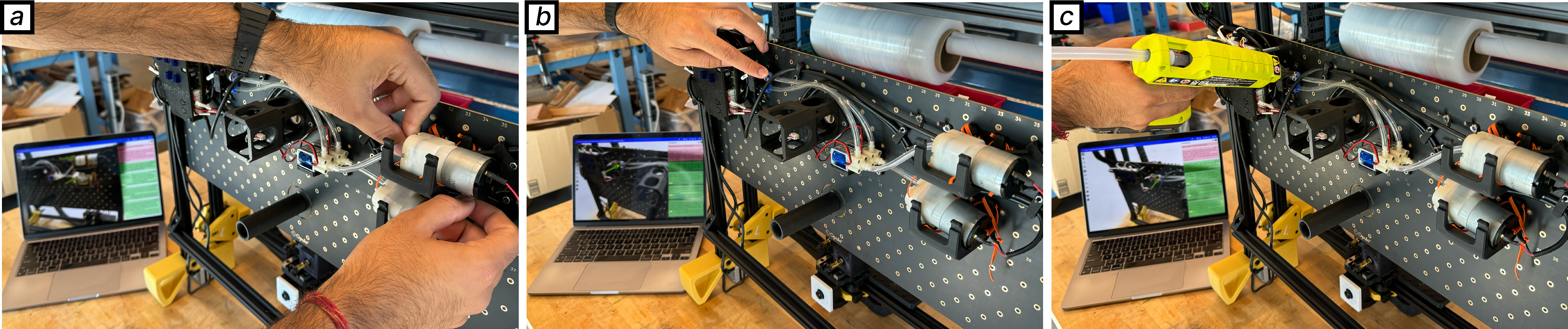}
    \caption{The local user examines the instructions left by the remote maintainer. (a) They test whether the tubing is still firmly attached to the vacuum pump. They feel that it is, so they move to the next suggestion. (b) They test whether they can feel any air escaping the fitting that the maintainer referenced in their suggestion. They can feel air leaking out. (c) The local user tries the stop-gap solution of applying some hot glue to the fitting to improve the seal.}
    \Description{
    Figure 8. This figure has three subfigures in it: A, B, and C. The overall figure shows a local user following the asynchronous instructions of a remote maintainer. The right subfigure A shows the local use testing whether the tubing on the machine is still firmly attached to the vacuum pump. The middle subfigure B shows the local user testing whether they can feel any air escaping the fitting that the remote maintainer referenced. The right subfigure C shows the local user trying the stop-gap solution of applying some hot glue to the fitting to improve the seal.
    }
    \label{fig:so-trying-suggestions}
\end{figure*}

The local user reviews the suggestion and performs the test as described by the maintainer, as shown in Figure \ref{fig:so-trying-suggestions}. When they turn on the pump, they feel air escaping the fitting. They apply some hot glue, and the vacuum pressure becomes more substantial. The local user orders a replacement fitting and replies to the suggestion indicating that a poor fitting was the culprit. Once the issue is resolved, the timeline is indexed to all the parts referenced in the back-and-forth exchange.

% \color{purple} % start CHANGE
\section{Utility of SplatOverflow} \label{sec:util}
This section outlines the capabilities made possible by SplatOverflow. 
We discuss the scenarios in which SplatOverflow addresses the limitations of existing troubleshooting artifacts, highlight the novel capabilities SplatOverflow enables, and outline the utility it creates for different users.
\subsection{Existing Troubleshooting Artifacts}
Existing hardware troubleshooting workflows use a variety of artifacts to facilitate back-and-forth communication between parties, each with its own drawbacks. 
We highlight how SplatOverflow mitigates these drawbacks by complementing each kind of artifact with additional information and context within a scene.

\subsubsection{Text Posts}
Text is commonly used to describe the physical state of hardware, how it may be erroring, or what action must be taken on the hardware.
However, text descriptions can create confusion if parties do not share a common vocabulary. This is especially true when non-expert users are troubleshooting hardware with maintainers.
Non-expert users may not know how to refer to specific parts or which part a provided instruction refers to. 
SplatOverflow explicitly links text to the referent parts from the CAD model and visually highlights corresponding parts when the text is selected. 
This preserves the flexibility of text while addressing issues of referential uncertainty and ambiguity.

\subsubsection{Annotated Images}
Annotated images direct attention to a specific area or part of the hardware. 
Moreover, annotations allow maintainers to specify instructions through deictic references on the local user's hardware, e.g., tighten this bolt (where \textit{this bolt} is circled in an image).
This lets local users reason about instructions without changing contexts, as the annotation is authored by referencing their hardware. 
However, image annotations can only reference what the local user captured in the image. 
When the image does not capture the appropriate details, maintainers must guide the user without context cues from the local user's hardware. 
SplatOverflow extends the utility of image annotation for maintainers by allowing them to annotate the scene from various arbitrary viewpoints. 

\subsubsection{Video Tutorials}
Video tutorials can capture detailed multi-step processes on hardware and show local users how to manipulate their hardware through demonstration. 
Video tutorials contain a large amount of \textit{detail}, but can demand significant time and planning to be effective. 
As a result, video documentation is often used for build instructions or set-up guides, the utility of which can be pre-empted and planned for. 
However, troubleshooting is often about problems that are hard to preempt \cite{orr-talking-about-machines}. 
Creating detailed video tutorials for each of these issues as they arise is demanding for maintainers.
In contrast, impromptu and unedited videos can be difficult for users to parse, leading to more miscommunication.
SplatOverflow improves the utility of \textit{impromptu video} filmed by non-expert users by situating the filming perspective of the video within the scene. 

\subsection{Novel Capabilities}
In addition to addressing some of the limitations of existing artifacts, SplatOverflow enables novel capabilities that are not supported by existing asynchronous hardware troubleshooting artifacts.

\subsubsection{Indexing Issues}
In SplatOverflow, resolved issues are indexed in a database based on the parts they reference. Future users can subsequently query this database by selecting the parts they are experiencing issues with. This streamlines accessing the evolving technical information related to the hardware.
Similar to the pools of technical knowledge created by the social network of copier repair technicians described by Orr \cite{orr-talking-about-machines}, the indexed corpus of hardware issues created by SplatOverflow is a valuable reference for other maintainers and future users \cite{bobrow_community_2002}.

\subsubsection{Independent Asynchronous Navigation}
With SplatOverflow, maintainers can independently navigate and author instructions from perspectives and on parts not intentionally captured by the local user. 
This adds flexibility and enables maintainers to inspect and investigate the actual state of the local user's hardware.

\subsubsection{Issue Re-contextualization}
By defining annotations in reference to CAD data, SplatOverflow can re-contextualize them into new scenes containing the same hardware. 
This allows local users to view suggestions from past issues overlaid onto their own hardware and reduces the amount of context shifting needed to make sense of instructions from past issues. 

\subsection{Utility to Users}
SplatOverflow mediates an interaction between two kinds of users, non-expert \textit{local} end-users and expert \textit{remote} maintainers. This section underscores the utility to each type of user and the conditions in which SplatOverflow is most beneficial.

\subsubsection{For End-Users}
SplatOverflow provides the most utility to users whose expertise is not in the hardware itself but in what they use it for. 
These users depend on the hardware for productivity but are not familiar with the intricacies of its design. 
As such, maintenance and troubleshooting are simultaneously critical and challenging tasks for these users to carry out. 
SplatOverflow benefits these end-users in three ways:
(1)~By indexing past issues and relevant technical information, SplatOverflow helps end-users search for workflows to remedy issues they are facing.
(2)~By re-contextualizing past issues, SplatOverflow allows end-users to reason about and follow existing instructions within the context of their own hardware.
(3)~By capturing the hardware context in a scan, SplatOverflow allows end-users to seek help without knowing hardware-specific terminology.
These benefits are accessed by capturing a brief video of their hardware.

\subsubsection{For Maintainers}
As hardware develops, end-user maintenance and troubleshooting is essential to support \cite{dunn_jubilee_2023, orr-talking-about-machines, hodges_long_2019, eghbal2020working}.
This entails hardware designers, producers, or enthusiasts taking on the role of maintainers to support the growing number of non-expert end-users.
SplatOverflow helps maintainers scale support for hardware in three ways. 
(1)~By creating an asynchronous troubleshooting workflow, SplatOverflow allows maintainers to address end-user issues at their own pace.
(2)~By registering an immersive scan to a CAD model, SplatOverflow allows maintainers to compare the actual state of the hardware to its design without context shifting across artifacts.
(3)~By indexing resolved issues, SplatOverflow allows maintainers to inspect and share links to related issues and relevant solutions. 

These features help maintainers attend to individual issues. However, as more issues are addressed and indexed through SplatOverflow, the benefits to maintainers expand. For example, the maintainer's effort in troubleshooting issues is multiplied when future end-users can reference the solution without seeking out help. Moreover, when multiple people are responsible for maintenance and troubleshooting, they can reference the pool of advice offered by others in the past while diagnosing new issues.
As such, SplatOverflow supports maintainers in scaling hardware support.
\section{Related Work}
SplatOverflow builds on studies of hardware maintenance practices, documentation systems, telepresence systems for expert support, and spatial rendering techniques. 

\subsection{Hardware Maintenance Practice}

SplatOverflow draws on the findings of Julian Orr's ethnographic work, which studies the sharing practices of Xerox repair technicians in the field \cite{orr-talking-about-machines}. Orr's work found that such troubleshooting was improvisational and centered collaborative sensemaking between a technician, a client, and a machine \cite{orr-talking-about-machines}. 
This form of collaboration between different parties remains essential for effective maintenance in modern, distributed communities of hardware users. \citet{dunn_jubilee_2023} study one such distributed community: users of the Jubilee motion platform, an open-source hardware project. They find that encouraging community-driven maintenance enabled the hardware to evolve to address design issues that would likely have gone unaddressed otherwise \cite{dunn_jubilee_2023}. Orr emphasizes how hardware maintenance and troubleshooting practice cannot exhaustively be preempted through documentation \cite{orr-talking-about-machines}. The practical nature of hardware issues is that they are non-canonical and can only be addressed through pools of shared knowledge that continue to evolve over time \cite{bobrow_community_2002, Brand2025-hc}. 

More recently,~\citet{subbaraman-maintenance} argue that maintenance is a core component of using digital fabrication workflows in 3D printing communities. Like Orr, they caution against addressing maintenance needs with fully automated procedures and instead advocate for systems to be designed with maintenance in mind.
This resembles Jackson's \cite{jackson_rethinking_2014} concept of "broken world thinking," which argues for foregrounding repair and maintenance practices when considering technological progress. SplatOverflow revisits these ideas to examine how such troubleshooting could be coordinated asynchronously and how systems supporting hardware maintenance can contribute workflows to pools of practical knowledge in hardware communities.

% We aim to synthesize findings from earlier work into a technical system to support hardware troubleshooting and establish infrastructure for scaling maintenance practices for hardware.

\subsection{Documentation Systems}
Documentation for assembly instructions, usage guides, and maintenance tasks \change{is} integral to how users interact with and navigate hardware. Such documentation is distinct from the \change{design} files that define the hardware, but researchers have argued that they contribute just as significantly \change{to its success and adoption} ~\cite{bonvoisin_what_2017, orr-talking-about-machines}. 
To this end, HCI researchers have examined how hardware designers can efficiently create documentation that is useful to end-users. ~\citet{mariscal-melgar_semi-automatic_2023} propose a semi-automated method for generating and updating assembly documentation by leveraging data from the hardware's CAD model. This lets designers compile new assembly instructions as designs change and keep documentation in sync with updates to the hardware.
~\citet{milara_document-while-doing_2019} propose software tools for makers to create documentation as they work on projects to ensure that essential design decisions, lessons from failures, and the process narrative are not lost. 
\citet{tran_oleary_tandem_2024} demonstrate how to create fabrication workflows in a manner that foregrounds reproducibility by others. By interweaving digitally controlled processes and physical interventions when defining a fabrication workflow, \textit{Tandem} guides users through the manual steps critical to successfully reproducing a workflow \cite{tran_oleary_tandem_2024}. 

Researchers have also explored how to make it easier for end-users to explore existing technical documentation. For instance, \textit{MagicNeRFLens}~\cite{li_magic_2023} allows users to navigate a NeRF scene in virtual reality, with additional documentation overlaid onto the virtual scene. \textit{ARDW}~\cite{chatterjee_ardw_2022} is an augmented reality projection workbench that overlays documentation onto a PCB to support debugging. These projects present overlay and interaction techniques that address challenges associated with context shifting between design, documentation, and physical instantiation of hardware. This research underscores that the successful documentation, distribution, and reproduction of hardware and hardware workflows extends far beyond sharing digital design files. However, these projects demonstrate that there is significant value in incorporating connections between digital design files, technical documentation, and physical instantiations of hardware. SplatOverflow builds on this work by leveraging existing CAD model information to author, share, visualize, and index hardware troubleshooting workflows.

\color{black} % end CHANGE
\subsection{Telepresence for Expert Support}

In software development~\cite{anderson_discovering_2012} and document editing~\cite{posner1993people}, all collaborators typically access the same source code instead of compiled or high-level artifacts. Direct access to the collaborative artifact enables asynchronous support across the web~\cite{anderson_discovering_2012}.

HCI researchers examining synchronous support have proposed multiple theories and developed speculative systems for how remote experts can virtually occupy a shared task space with non-expert users to provide guidance~\cite{buxton1992telepresence, feiner-virtualreplicas, gauglitz_integrating_2012, fussell-coordination}. 
Telepresence solutions often try to establish a shared reference space, in which participants can communicate with one another with deictic expressions and gestures as they would if they were in person~\cite{johnson-know-where-that-is, gutwin_descriptive_2002, dourish_awareness_1992}. Such a reference space depends on both participants being able to communicate within a common context.  In a survey on augmented reality systems for collocated experiences,~\citet{needs-survey-AR-collab-cscw} found that collaborators need a shared environment to ground their communication. Moreover, they find that collaborators need tools to guide attention and give instructions by referencing objects in the shared task space.~\citet{chastine-inter-referential} develop a formal model of such inter-referential awareness, providing a detailed framework for how designers can develop referential systems.

Advances in augmented and virtual reality have led to work studying how experts can remotely collaborate on physical tasks~\cite{wang2021ar}. Researchers have proposed systems for 'mentoring' tasks, where one less experienced user seeks out guidance from an expert user who views the scene remotely using mobile AR to bring the physical environment of users to remote collaboration~\cite{fussell_coordination_2000,johnson-know-where-that-is, gauglitz_integrating_2012}. Some systems let experts view the perspective of remote users in VR and communicate guidance through overlaid annotations~\cite{transceiVR-VR+externals, collab-pointing-ar, yu2022duplicated}. 

Alternatively, systems also capture 3D reconstructions~\cite{izadi_kinectfusion_2011} and allow remote collaborators to interact by combining with teleoperated robots as in \textit{VRoxy}~\cite{sakashita_vroxy_2023} or by integrating live 360\textdegree video as \citet{Theophilus-360video-3Dreconstruction} demonstrated.

In these systems, the interaction is synchronous, and the guidance provided is ephemeral, making it difficult for third parties to replay and observe at a later time. \textit{Heimdall}~\cite{Heimdall} is an alternative approach to remote collaboration for prototyping electrical circuits featuring a bespoke scanning system. Instructors navigate around the prototype circuit and inspect its schematics through an instrumented breadboard. This enables instructors to remotely examine and debug student circuits without relying on a student to navigate around their breadboard. \change{SplatOverflow marries insights from synchronous remote expert support systems with the flexibility and observability enabled by asynchronous troubleshooting.}

% \color{purple} % start CHANGE 
\subsection{Scanning and Spatial Rendering}

To troubleshoot hardware issues, SplatOverflow must be able to capture a local user's hardware, ideally without specialized equipment. To do this, it leverages research in scanning and scene representation techniques. Recently, there has been substantial progress in novel-view synthesis techniques that can construct a 3D scene from collections of posed 2D images ~\cite{mildenhall_nerf_2020, muller_instant_2022, kerbl_3d_2023}. These approaches build on COLMAP, a structure-from-motion (SfM) tool that constructs a coarse point cloud and estimates camera pose from image frames ~\cite{snavely_photo_2006, schoenberger2016sfm, schoenberger2016mvs}. These tools and techniques allow SplatOverflow users to capture high-fidelity scans using commodity smartphone cameras they likely already own. 

Although SplatOverflow is generally agnostic to scanning technology, our implementation leverages 3D Gaussian Splatting \cite{kerbl_3d_2023} to capture a workspace due to its fast training time and real-time rendering capabilities. A splat represents the scene as a collection of 3D Gaussian distributions and can be rendered in real-time using traditional graphics pipelines. Notably, there is considerable progress in reducing training times and improving the visual definition of Gaussian splats \cite{guedon_sugar_2023, liu_mvsgaussian_2024, wang_dust3r_nodate}. These methods enable fast 3D scene reconstruction with accurate geometric information from limited data, lowering the overhead of capturing a scene for users. 
\section{SplatOverflow}
% \begin{figure}
%     \centering
%     \includegraphics[width=\linewidth]{figures/splatoverflow-components.pdf}
%     \caption{The components of SplatOverflow's interface. (a) is the palette of gestures SplatOverflow offers for selecting parts, guiding attention, and communicating actions. (b) the 3D Gaussian Splat of the hardware is aligned and registered onto the CAD model. (c) the timeline that captures the troubleshooting interaction as a sequence of instructions and responses between users.}
%     \label{fig:so-ui}
% \end{figure}

This section describes the design of SplatOverflow. We discuss the constituent artifacts that comprise a SplatOverflow scene, including the benefits and shortcomings of each artifact. Then, we explain the construction of the SplatOverflow scenes, how it facilitates asynchronous communication using \textit{gestures}, and how data from issues can be indexed and retrieved using a CAD model. 

\subsection{Constituent Artifacts}
At a minimum, SplatOverflow requires a scan of the hardware and a CAD model that describes the hardware's design. SplatOverflow combines these artifacts to construct a single boundary object that amplifies the benefits and minimizes the drawbacks of each constituent artifact, as well as addressing the limitations of traditional asynchronous troubleshooting artifacts described in Section \ref{sec:util}.

\subsubsection{CAD Models}
CAD models can precisely and accurately describe the design of hardware. They also offer various viewing, selection, and manipulation tools to help make sense of hardware issues. For example, CAD models allow designers to inspect internal mechanisms that are \change{not} visible in the assembled hardware. This ability to peer through parts can help designers reason about mechanical issues without taking apart the hardware. These models can also be the basis of generating visual assembly instructions that are easy for users to follow \cite{agrawala-assembly-instructions, mariscal-melgar_semi-automatic_2023}. 

The primary drawback of CAD models is that they do not reflect the host of ways hardware can err. This is because CAD models capture the hardware in an idealized state. In practice, however, hardware (especially malfunctioning hardware) does not perfectly resemble the CAD model. This is additionally problematic as some deviations from the CAD model are perfectly acceptable, while others can introduce errors. CAD models offer the tools and interactions to visualize, explore, and reason about a hardware design but do not capture the multitude of ways that each hardware instance is unique.

\subsubsection{Scans}
Scans, however, can capture the unique details of hardware instances. Dedicated mobile scanners using stereo vision, structured light, or novel-view synthesis techniques can accurately capture the geometry of hardware at different scales. This data can be used to measure and inspect hardware as it has been built and assembled, which is essential when diagnosing what may be going \textit{wrong} with the hardware. 

The drawbacks of relying on scans are that they only capture visible geometry and do not know the semantic meaning of what they are capturing. Most scanning methods cannot safely penetrate materials to scan internal components, so scans capture only the outer shell of the hardware geometry. Moreover, as scanning methods often do not know what they are scanning, isolating different parts in an assembly is challenging. These drawbacks are related in that they highlight how references are hard to anchor to scan data.

% \subsubsection{Video}

\subsection{A SplatOverflow Scene}
SplatOverflow scenes comprise two essential artifacts: a scan of the user's workspace aligned and registered onto a CAD model of the hardware. Our implementation generates the scans using 3D Gaussian Splatting \cite{kerbl_3d_2023}. We chose Gaussian Splatting as it can generate high-quality scans rapidly using only user-captured video data as input. We use open-source CAD models saved as .glb files with additional data to capture the assembly constraints in the model. The 3D Gaussian Splat (referred to here as splat) is aligned and registered to the CAD model using ArUco markers that are placed on the hardware at known locations \cite{munoz2012aruco, olson_apriltag_2011}. 

The aligned artifacts support selecting and manipulating individual components on the hardware segmented using the CAD model and inspecting the physical state of the hardware as captured in the splat.  SplatOverflow scenes are implemented in WebGL and run in the browser. As a result, they can be shared with anyone online, requiring no installation or OS-specific set-up. Scenes can be viewed on personal computers, mobile devices, and virtual reality headsets that support WebXR.

\subsubsection{Scanning a Workspace}
Scanning the hardware and surrounding workspace captures the as-built state of the hardware. This includes modifications, job configurations, or parts not traditionally included in CAD, such as wires or cables. 

The input to the scan is a video of the user's workspace \change{captured through SplatOverflow's mobile interface, as described in Section \ref{sec:walkthrough}}. The video is then sampled in two passes. First, at a fixed frame rate (4 FPS in our examples) and then again to more densely sample frames containing ArUco tags \change{to generate a set of images}. SplatOverflow feeds this set of images to COLMAP \cite{schoenberger2016sfm, schoenberger2016mvs} to generate an intermediate Structure-from-Motion model of the local users' workspace. SplatOverflow uses \change{this model} to generate a splat of the recorded scene.

SplatOverflow takes roughly four minutes to generate a 3D Gaussian Splat from a video filmed in 1080p resolution \change{and roughly two minutes from a 360p video. Generating the splat is the primary bottleneck for developing a SplatOverflow scene. Processing times could be reduced by using more efficient splat generation techniques from recent research, such as MVSGaussian ~\cite{liu_mvsgaussian_2024}.} 

% As described in Section \ref{sec:walkthrough}, SplatOverflow first generates a splat using the downsampled 720p video to develop a scene for the user quickly and replaces the splat with one created using the 1080p video when it is done generating.}

\subsubsection{Registering a Gaussian Splat to a 3D CAD Model}
SplatOverflow \change{automatically} aligns and registers a splat of the local user's hardware to the 3D CAD model in two steps using ArUco tags placed in known locations on the hardware \cite{munoz2012aruco}. SplatOverflow first uses the SfM model generated by COLMAP \cite{schoenberger2016mvs, schoenberger2016sfm} to compute a three-dimensional coordinate of every corner of each ArUco tag \cite{meyer_aruco_2023}. These corners are used to rescale the splat to be the same size as the CAD model. Next, SplatOverflow uses Arun's method \cite{arun_least-squares_1987} to compute a rigid transformation going from corners in the SfM model's coordinate space to CAD coordinate space. If there are no moving parts in the CAD assembly, this transformation yields an aligned and registered splat. 

However, hardware will often have moving parts. These degrees of freedom mean that the hardware can be in one of many states, which are unlikely to match the static state of the CAD model. To address this,  SplatOverflow recommends placing at least one \textit{constraint tag} on each degree of freedom and one \textit{grounding tag} on a stationary part of the hardware. When a CAD model is being prepared, the model's maintainer indicates which tag ID maps to which degree of freedom. With this precondition, SplatOverflow uses the same detection method to find the centers of each \textit{constraint tag} and computes an offset from the \textit{grounding tag} to the \textit{constraint tag}. This offset is then used to calculate a transformation that automatically aligns the CAD component to the position of the part in real life. Following these two steps, the hardware parts are aligned to their corresponding parts in the CAD model. 

The tags used for alignment, registration, and constraint satisfaction can be applied manually onto existing hardware, placed during manufacturing, or designed into the hardware \cite{dogan_infraredtags_2022, dogan_structcode_2023}. In all of our examples, tags were manually placed onto existing hardware.

\subsubsection{Post Processing a Splat Using Features of the CAD Model}
\begin{figure}
    % \centering
    \includegraphics[width=\linewidth]{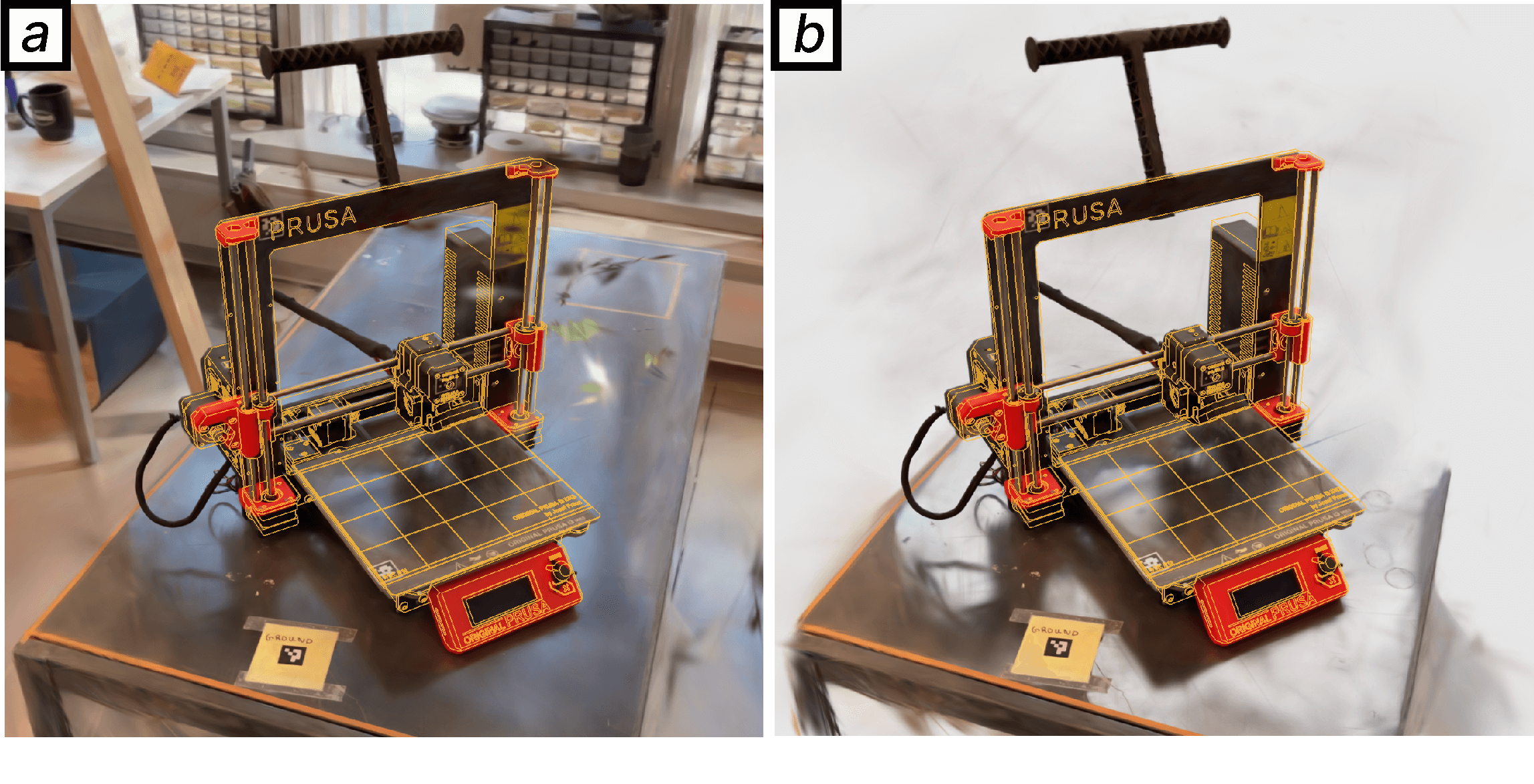}
    \caption{(a) An unaltered 3D Gaussian Splat as returned by our system. (b) A pruned 3D Gaussian splat is used to preserve the user's privacy and not share background details with a community of users.}
    \Description{
    Figure 9. This figure has two subfigures in it: A and B. The left subfigure A shows an unaltered Gaussian splat of a 3D printer used by SplatOverflow. The right subfigure B shows a pruned Gaussian splat of the 3D printer that is used to preserve the user’s privacy and not share background details with a community of users.
    }
    \label{fig:background-removal}
\end{figure}

Once the splat has been aligned and registered, SplatOverflow leverages information from the CAD model to post-process the splat. SplatOverflow uses a signed distance function from points in the splat to CAD meshes \change{to identify what region of the splat contains relevant information about the hardware}. \change{By default, SplatOverflow uses this to remove the background details from a scene to preserve the local user's privacy and share only parts of the splat that correspond to the hardware or the workspace the hardware rests on}. Figure \ref{fig:background-removal} shows a splat as it is generated and the same splat after SplatOverflow's privacy filter.
\subsubsection{Localizing Video Feeds}
SplatOverflow allows users to capture dynamic details \change{of their issue} as short video clips and aligns these clips in the SplatOverflow scene using an estimate of the camera pose at each frame. This allows remote maintainers to review the perspective from which a video was shot as well as the content of the video. Figure \ref{fig:so-video-embed} shows the video feed in the SplatOverflow scene. SplatOverflow estimates the pose of a video feed by localizing \change{individual} frames into an existing COLMAP model \cite{schoenberger2016sfm, schoenberger2016mvs}.

\subsection{Interacting with a SplatOverflow Scene}
SplatOverflow offers tools to explore a scene, communicate actions, and facilitate discussion.
This section describes each of these tools and the interactions they support. 

\begin{figure}[ht]
  \includegraphics[width=\linewidth]{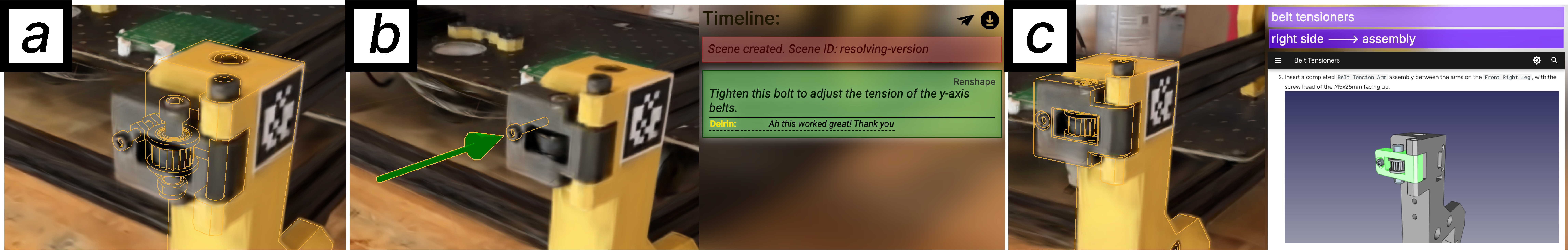}
  \caption{(a)~ \change{A} sub-assembly on the Lumen v3 with occluded components visible to the user. (b)~ \change{A} gesture indicating that the user tightens a bolt on the sub-assembly. (c)~ \change{Q}uerying technical documentation referencing a specific sub-assembly.}
  \Description{
  Figure 10. This figure has three boxes in it: A, B, and C. The left subfigure A shows the outlines of the internal components of a subassembly being rendered in the splat. The middle subfigure B shows an arrow point towards a bolt in a subassembly, indicating the bolt be tightened. The right subfigure C shows the user querying documentation on how to tension the belts on the right side of the machine.
  }
  \label{fig:interaction-examples}
\end{figure}

\subsubsection{Selection and View Control}
SplatOverflow allows users to select parts and navigate subassemblies of the CAD model to compare the as-designed to the as-built and make sense of occluded parts of the hardware not visible in the splat.
\change{When a part from the CAD model is selected, it is rendered to the user with a yellow wireframe. This allows the user to compare details from the splat to the geometry of an individual CAD component.}
\change{Moreover, the visibility of CAD parts in a subassembly can be controlled, allowing} users to "peer through" sub-assemblies and \change{visualize} internal components that are \change{not} visible, as shown in Figure \ref{fig:interaction-examples}\change{(a)}. 

Remote users can use SplatOverflow scenes to independently explore novel viewpoints within the local user's environment. This is enabled by the splat component of a SplatOverflow scene. Once the splat is aligned onto the CAD model, it occupies the same coordinate space and can be navigated with the same controls as a CAD environment.

\subsubsection{Gestures}
Users communicate in SplatOverflow via \textit{gestures}. \change{A gesture is an instruction or request that references an element in the SplatOverflow scene.} Gestures are organized chronologically in an interactive timeline. The timeline elements trigger visualizations of the gesture in the SplatOverflow scene and house user discussions about the gesture.

\change{Maintainers can use gestures to request more information from the user.} This can be done by requesting video annotations or updated splats using the \textit{request} gesture. When a request gesture is authored, a timeline element appears with a QR code. The QR code lets the user populate the request via SplatOverflow's mobile interface \change{and appends the requested information to the scene}. Figure \ref{fig:so-reorienting} shows an example of the request gesture.

Gestures can also communicate instructions or actions a user should take on their hardware. These gestures are all anchored to parts in the CAD model, which are highlighted whenever a gesture is selected in the timeline alongside each gesture's animation. SpaltOverflow's \textit{pointing} gestures indicate where on a part in CAD a user should inspect \change{or attend to}. When selected, the referenced CAD part is highlighted, and the viewport shifts to re-orient the local user. SplatOverflow's \textit{move} gesture indicates to local users how parts should be manipulated. When selected, the referenced CAD part animates to tween between its original position and the final position defined by the remote maintainer. \change{Finally, SplatOverflow includes a set of action gestures for common troubleshooting operations, such as tightening or loosening specific bolts and probing particular pads on a PCB. When selected, the gestures overlay arrows onto the scene to visualize the operation.}

\subsubsection{Timeline Elements and Discussions}
Timeline elements refer to gestures authored by different users. The timeline element contains additional text to clarify what a \change{gesture} entails. Users can ask for clarification or follow up on a gesture by replying to the corresponding element in the timeline. This facilitates the asynchronous back-and-forth communication to help make sense of a given gesture.

\subsection{Indexing Data to a CAD Model}
SplatOverflow uses the CAD model to index and query data. This includes technical documentation referencing parts in the CAD model and past SplatOverflow issues. For existing technical documentation, SplatOverflow renders web pages that reference CAD components, as illustrated in Figure \ref{fig:interaction-examples}(c). Similarly, SplatOverflow can query the history of past issues to retrieve troubleshooting instructions that reference specific CAD components. This ability to feed structured data from troubleshooting issues into a database that can be queried allows SplatOverflow to become increasingly useful to hardware communities over time.

\section{Implementation}

% \begin{figure*}[ht]
% \includegraphics[\textwidth]{figures/system.png}
%     \caption{A flowchart illustrating our implementation of SplatOverflow and the technologies it builds on.}
%     \label{fig:system-diagram}
% \end{figure*}

% \begin{figure}
% \includegraphics[width=0.5\textwidth]{figures/system.png}
%     \caption{A workflow illustrating our SplatOverflow implementation and the technologies it builds on.}
%     \label{fig:system-diagram}
% \end{figure}

% As shown in Figure~\ref{fig:system-diagram},
Our implementation of SplatOverflow is divided into four sections: Scene Capture,  Scene Generation, the SplatOverflow front-end, and the back-end Coordination Server. 

The mobile capture interface runs in a browser and is written in JavaScript using the MediaStream Recording API. It is designed to be accessible from modern smartphone browsers. On the server, SplatOverflow uses a C++ 3D Gaussian Splatting implementation by MrNeRF~\cite{kerbl_3d_2023} and the COLMAP package for Structure-from-Motion (SfM) to generate the scene. To align a CAD model to the scan, SplatOverflow uses the method from ArUco SFM Scale Adjustment~\cite{meyer_aruco_2023} and an implementation of Arun's method \cite{arun_least-squares_1987}.

The back-end coordination server is a Flask application and SQL database that manages the data associated with a SplatOverflow scene. The SfM \cite{schoenberger2016sfm} and Gaussian Splatting \cite{kerbl_3d_2023} pipelines run on a PC equipped with an Intel i7-14700K, 32GB RAM DDR5 and Nvidia 4080S GPU with 24GB of VRAM. On average, a 60-second video takes 122 seconds to generate a scene trained on 360p footage and 250 seconds to generate a scene trained on 1080p footage.

We use Mark Kellogg's 3D Gaussian Splatting Renderer~\cite{kellog_3d_2024} to render the 3D Gaussian Splats and develop the interface elements in Javascript, using the Three.js library~\cite{danchilla_threejs_2012}. 
\section{Demonstrative Examples} \label{sec:demo}

\begin{figure*}[t]
  \includegraphics[width=\textwidth]{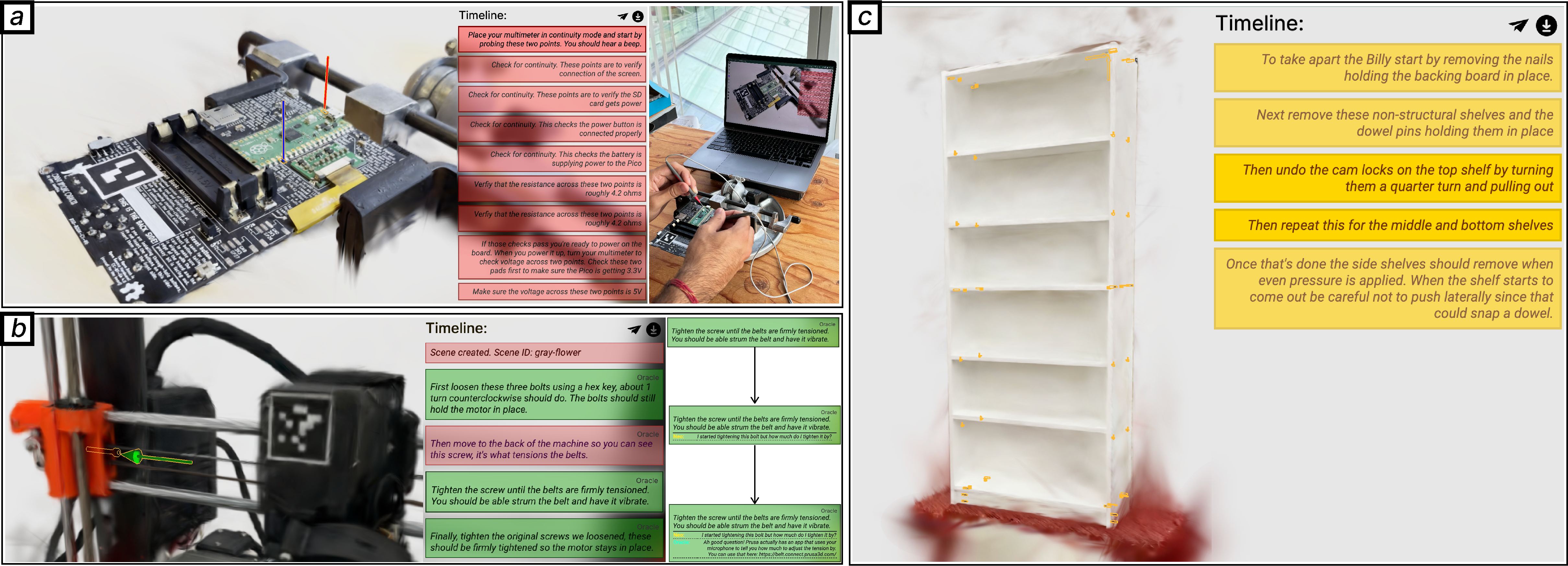}
  \caption{\change{SplatOverflow functions with hardware of varying scales}. (a)~ \change{Assembly verification for Printed Circuit Boards.} (b)~ \change{Community authored troubleshooting best practices}. (c)~ \change{Disassembly process for mass-produced flatpack furniture.}}
  \Description{
  Figure 11. This figure has three boxes in it: A, B, and C. The left upper subfigure A shows a PCB with a sequence of instructions showing where a user should probe to validate their assembly and an image of a user performing the probing. The left lower subfigure B shows a workflow to adjust the tension of the belts on a 3D printer. It also shows a sequence of replies that add information to the original instruction. The right figure C shows an ikea bookshelf alongside instructions on how to disassemble it.
  }
  \label{fig:examples}
\end{figure*}

Figure ~\ref{fig:examples} demonstrates the utility of SplatOverflow in three workflows that extend beyond troubleshooting of machines: (a)~verifying assembly of the PCB in the \textit{Open Book} e-reader by Oddly Specific Objects, (b)~ performing routine maintenance on the \textit{Prusa MK3S} 3D printer, and (c)~following disassembly instructions for a flat-pack bookshelf using the \textit{Billy} bookshelf by IKEA. These workflows intend to capture a variety of scales, applications, and user roles that SplatOverflow can support.

% \begin{figure}[ht]
%   \includegraphics[width=\linewidth]{figures/demon-examples.pdf}
%   \caption{Other workflows that extend the utility of SplatOverflow (a)~ non-mechanical CAD models (b)~ asynchronous workflows to create feedback loops on instruction from the community (c)~ enabling workflows on existing furniture with a simple sticker.}
%   \Description{}
%   \label{fig:examples}
% \end{figure}

\subsection{Verifying the Correct Assembly of the \textit{Open Book} E-Reader}
SplatOverflow is compatible with non-mechanical CAD designs. 
The \textit{Open Book}~\cite{castillo_open_nodate} e-reader is an open-source e-reader that consumers assemble on their own, with the goal of demystifying consumer electronics. Successful assembly is a function of correct and precise soldering. In this example, we show how the maintainer of the Open Book can help users verify correct assembly and track down potential errors that inhibit successful booting using a workflow authored and shared in SplatOverflow. Specifically, the maintainer defines a set of probe points to be checked with a multimeter that can indicate the exact probe points on the PCB the user must test, along with the expected outcome of each test.

% \begin{figure}[ht]
%   \includegraphics[width=\linewidth]{figures/probing-guide.pdf}
%   \caption{(a)~ The assembly verification guide overlaid onto a splat of the local user's assembled Open Book. (b)~ The local user follows instructions to validate their assembly.}
%   \Description{}
%   \label{fig:probing-guide}
% \end{figure}

The local user starts by creating a SplatOverflow scene to follow the assembly verification guide created by a maintainer. The Open Book has a 20mm tag that aligns the CAD model of the PCB generated from KiCAD~\cite{kicad}. In this example, we applied the tag to the PCB, but in practice, such a tag could be printed as part of the PCB's silkscreen. Once the SplatOverflow scene is generated, the local user loads the instructions authored by the maintainer into their scene and sees them projected onto their hardware. Figure ~\ref{fig:examples}(a) shows the instructions rendered onto their PCB. 

The local user references verification instructions loaded into their SplatOverflow scene. As shown in Figure ~\ref{fig:examples}(a), instructions are overlaid onto the pins the user must probe. The user follows probing instructions to check for continuity, voltage, and resistance at points across the PCB. The guide authored in a SplatOverflow scene can be \change{recontextualized} to each local user's hardware. This contrasts most online guides, which are rendered using virtual objects or images and videos of a different hardware instance. 

% The user follows probing instructions to check for continuity, voltage and resistance at points across the PCB, as shown in Figure ~\ref{fig:examples}(b). At each step, the timeline event indicates the expected result of the probing for the user to compare against. The guide authored in a SplatOverflow scene in this application can be contextualized to each local user's hardware. This contrasts most online guides, which are rendered using virtual objects or images and videos of a different hardware instance. 

\subsection{Sharing and Updating Routine Maintenance Workflows on the \textit{Prusa MK3S} 3D Printer}
Asynchronous sharing enables feedback loops, leading to improved workflows.
The \textit{Prusa MK3S} 3D Printer is a popular personal fabrication machine targetting hobbyists. Like all 3D printers and fabrication machines, the printer requires routine maintenance to ensure proper function. A typical maintenance task is tightening the belts that drive two axes of motion. Poorly tightened belts can lead to failed prints or unintended artifacts in otherwise successful prints. Learning to carry out such tasks is essential to owning and operating personal fabrication machines \cite{subbaraman_3d_2023}. We illustrate how SplatOverflow can be used to share maintenance workflows and how end-user feedback can improve the quality of such workflows.

% \begin{figure}[ht]
%   \includegraphics[width=\linewidth]{figures/prusa-back-and-forth.pdf}
%   \caption{(a)~ A SplatOverflow scene rendering instructions to adjust the belt tension on the x-axis of the Prusa MK3S. (b)~ One of the instructions specifies how the belt tension is adjusted. (c)~ A comment from an end user asking for more clarification on an instruction. (d)~ A maintainer adding more information and linking to other online resources that end-users can reference.}
%   \Description{}
%   \label{fig:prusa-belt-tension}
% \end{figure}

A local user sees ghosting artifacts on their print and is pointed to a SplatOverflow issue explaining how to adjust the machine's belt tension. Figure \ref{fig:examples}(b) shows the workflow rendered in the local user's workspace. The user follows the instructions but is unsure how much to tighten the belts. They leave a comment asking the maintainer how they know when to stop tightening the belt. The maintainer realizes that the instruction is underspecified and shares an online resource that uses a microphone to validate belt tension. Here, we show a simple example of how the open and asynchronous nature of SplatOverflow issues allows communities of users to iterate on workflows together.

\subsection{Guided Disassembly of the \textit{Billy} Bookshelf}
SplatOverflow enhances static assemblies like furniture by adding a sticker.
The \textit{Billy} Bookshelf is a mass-produced piece of flat-pack furniture. IKEA provides DIY assembly through easy-to-understand assembly instructions. We demonstrate how SplatOverflow can guide users through step-by-step \textit{disassembly}. 

The \textit{Billy} is roughly 80 inches tall, making it difficult to scan from all angles; particularly from the top down. The local user places an ArUco marker on the back-right corner of the bookshelf and scans the \textit{Billy}. As the bookshelf is upright during the scan, there is a noticeable degradation of quality above the bookshelf. However, because SplatOverflow aligns the CAD model in the scene, users can infer geometry while still receiving context cues from the scan. 

The disassembly workflow is reasonably straightforward. The maintainer (in this case, IKEA) provides a SplatOverflow scene with the correct sequence of opening the cam locks that hold this bookshelf together. Users follow these instructions and sequentially disassemble the shelf. For more complex disassembly, users can update the splats to receive new instructions, similar to the scene update described in the walkthrough. This example demonstrates \change{how SplatOverflow can enable new collaborative workflows for mass-produced hardware}.

% This example shows relevance beyond hobbyist hardware and how adding a sticker to ordinary furniture can enable new collaborative workflows for everyday objects.

\section{Evaluation} \label{sec:eval}
We conducted a usability study with twelve participants to validate whether end-users can use SplatOverflow to troubleshoot hardware issues. The hardware used in the study was a popular 3D printer: the \textit{Prusa MK3S}. 

\subsection{Study Design}
Our study is divided into two parts. First, we examine whether non-expert users can capture their hardware to create a SplatOverflow scene. Second, we present them with two common issues experienced by users of this hardware. These issues were sourced based on the prevalence of online guides addressing them. We examine whether users can successfully follow instructions through a SplatOverflow scene and assess the usability of our system. 

\subsection{Part 1: Generating a SplatOverflow Scene}

\begin{figure*}[ht]
  \includegraphics[width=\textwidth]{figures/figure-12.png}
  \caption{We conducted a usability study with 12 participants to evaluate whether they \change{could} generate a SplatOverflow scene and successfully fix the issue. \change{(a) Sample scenes generated by study participants. (b) A table showing each participant's outcomes in our study and the length of the recording used to generate the scene.}}
  \Description{
Figure 12. Has two subfigures in it A and B. The left subfigure A shows six renderings of SplatOverflow scenes. Each corresponds to a user-generated splat in a user study. The splats vary in quality with worst on the left and best on the right. The right subfigure B shows a table of study results. 
  }
  \label{fig:evaluation-tracking}
\end{figure*}

% <table of results>
Participants were provided a smartphone running SplatOverflow's capture interface as a web app. They were instructed to create a splat of the 3D printer in front of them by recording a video of the hardware from a variety of angles. All participants were able to generate a splat from their recording, and 83.3\% of participants could successfully generate a SplatOverflow scene. For 33.3\% of participants, the system generated a SplatOverflow scene, but the user-captured video lacked enough information to resolve hardware constraints—specifically, aligning the degrees of freedom in the hardware. Figure~\ref{fig:evaluation-tracking} shows examples of scenes generated by participants. After generating the scene, we demonstrated how participants could navigate our interface. They then completed a survey about the usability of the scanning interface.

We use the participant-generated SplatOverflow scene in part 2 of our study. If the user's input failed to generate a scene, they were given a placeholder scene featuring the same hardware.

\subsection{Part 2: Following Instructions in SplatOverflow}
% <table of results>
\begin{figure}
  \includegraphics[width=\linewidth]{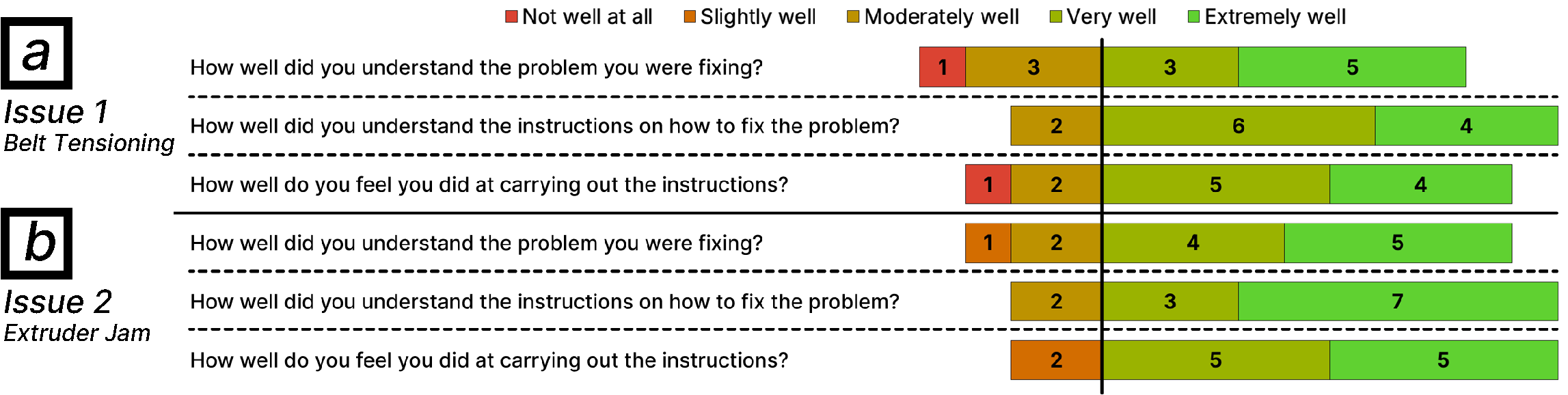}
  \caption{
  \change{Participants' responses to post-task questions regarding their confidence while troubleshooting the two isses.} (a)~ \change{Shows responses for }Issue 1: adjusting belt tension. (b)~ \change{Shows responses for }Issue 2: clearing an extruder jam.}
  \Description{
  Figure 13. This figure is a chart displaying the results of a Likert scale question. The questions ask how well users understood the problem they were fixing, how well they understood the instructions on how to fix the problem and how well they feel they did at carrying out the instructions. The chart repeats these questions for two issues the users fixed. The chat indicates most people felt positively about all questions in both conditions. 
  }
  \label{fig:post-issue}
\end{figure}

In part 2, we presented the participant with two hardware issues in random order and asked them to follow instructions rendered in SplatOverflow to remedy the issues. The issues were: correcting the belt tension in the printer's x-axis (issue 1) and clearing a jam in the extruder (issue 2). These issues were selected as they are commonly occurring issues with this printer and have a variety of online tutorials dedicated to them ~\cite{extruderJam,beltTension}. The order of the issues was randomized, and participants were asked to follow the instructions to the best of their ability. For both issues, 91.7\% of participants could successfully follow instructions in SplatOverflow to remedy the problem. Following each issue, we asked participants to answer questions about their experience using SplatOverflow to troubleshoot the issue. Figure ~\ref{fig:post-issue} shows participants' responses on a Likert scale. Overall, users found that instructions rendered in SplatOverflow were easy to follow and well-aligned with the scanned hardware.

\subsection{System Usability}
After completing both parts of the study, participants evaluated the overall usability of the system through a survey. We use the System Usability Scale \cite{brooke_sus_nodate} and our system scored 86.25 / 100.

In addition, participants shared insights on how SplatOverflow could be expanded to assist with troubleshooting in other contexts. Participants felt that SplatOverflow would be helpful when facing issues with unfamiliar hardware: \textit{"It is very helpful for me to use and repair, especially when I'm not familiar with the machine itself"} (Participant 1). They also compared our system to existing methods of troubleshooting: "\textit{I was working with robots (Turtlebot 4 setup). Their documentation was too wordy, and it was hard to locate the solution from that. So, we had to post GitHub issues and wait for a response. It was sometime hard to explain them my specific problem. Of course, if there was a system that I could have scanned my problem and send them, it would have saved a lot of time"} 
 (Participant 3).
\section{Discussion}
We discuss the implementation and application of SplatOverflow. We first highlight aspects of the experience for different users and then discuss technical features and future implementation.

\subsection{SplatOverflow Users}
\subsubsection{Hardware Support for End-Users}
SplatOverflow offers a new way to collaborate on hardware. 
For workflows such as assembly, routine maintenance, operating instructions, and troubleshooting, SplatOverflow provides a novel way for users to receive guidance from collaborators or retrieve associated documentation. 
By connecting physical instances of hardware to their CAD model and associated documentation, end-users can inspect the 'source' of the hardware they are working with. This direct connection between the physical part and documentation provides a more direct way to navigate documentation that can exist in disparate locations.

Moreover, users can seek help through SplatOverflow, analogous to asking for help in a forum. 
In software, platforms such as GitHub~\cite{dabbish2012social} and StackOverflow~\cite{anderson_discovering_2012} have been essential for helping users learn about new software, troubleshoot issues, and scale maintenance efforts to support distributed projects. 
By supporting asynchronous communication about physical things, SplatOverflow presents a similar workflow for hardware.

\subsubsection{Hardware Support for Maintainers}
Through SplatOverflow, remote maintainers gain insight into the local user's workspace and the actual state of the hardware they are working on. 
Moreover, maintainers do not have to rely on the local user to move a camera precisely to inspect hardware. They can examine novel viewpoints independently and use gestures to guide the local user to move to the exact location.

This is a longstanding problem in collaborative workflows previously addressed with teleoperated robotic systems ~\cite{Heimdall,sakashita_vroxy_2023}. SplatOverflow achieves a similar outcome without the need for bespoke hardware or additional instrumentation of a space. This, paired with SplatOverflow's web-based interface, expands the number of users who can access remote support.

\subsubsection{Reducing Barriers to Distribute Hardware}
Tech support is a core component of a viable hardware product but requires significant effort for hardware companies to scale. Asynchronous support infrastructure could reduce this effort by (a) providing communities of users the means of querying past issues and examining solutions and (b) allowing maintainers to provide support in an open format without coordinating with end-users. This approach would support maintainers, producers, and hardware communities in scaling technical support for users. To better evaluate this, we plan to engage with open-source hardware producers and community maker spaces to study longer-term deployments of SplatOverflow.

\subsection{Extending Multi-media Workflows}
Many tasks start with the assumption that the user is seeking guidance and has a screen available in front of them. For example, tools such as Interactive HTML BOM~\cite{htmlBOM} allow users to correlate components and their placements visually. The tool \textit{assumes} users will be using a screen to aid in assembly and debugging.

Similarly, YouTube and other video hosting platforms~\cite{utz2022videos, kim2014crowdsourcing, bhatia2018interdiscursive} are used extensively for step-by-step instructions in cooking, assembly, and DIY Home repair. Documentation often has directly embedded video ~\cite[p. 121]{gibb2015building} and manufacturers such as IKEA~\cite{Wagstaff_2012} produced videos to demonstrate operations clearly. While these videos can be valuable resources, they are often expensive and time-consuming to create. As a result, video is not always the ideal choice for end-users to articulate issues or seek guidance.

Our work extends multi-media hardware support by allowing users to rapidly author and share workflows for acting upon hardware amongst themselves. With SplatOverflow, bidirectional communication becomes easier instead of being solely broadcast from experts or designers.

\subsection{Cross-Referencing,  Inspectability and Information Sharing}
SplatOverflow enables the ability to link across hardware and data. This interaction mode is similar to \textit{View Source} or Right Click to \textit{Inspect Element} that is available in browsers~\cite{viewsource}. These features were originally intended to help developers debug their own software~\cite{flailingmonkey2013, luthra2010firebug}. However, they rapidly became tools for education and exploration where ``Every single web page you visited contained the code showing you how it was created. The entire internet became a library of how-to guides on programming''~\cite{thompson2020coders}. While we leave the specific application of SplatOverflow in education contexts to future work, we believe that the ability to inspect, discuss, and cross-reference physical and digital representations is vital to sharing hardware designs and know-how. SplatOverflow provides a set of primitives to support a known challenge for commons-based peer production~\cite{benkler2002coase, benkler2006commons} of physical systems~\cite{subbaraman_3d_2023}, primarily that of scalable instruction, support, and collaboration.

\subsection{Technical Features}
After implementing and testing SplatOverflow in a variety of contexts, we distilled a set of features to support it as we move towards a full deployment of the system.

\subsubsection{Guided Splat Capture}
In our evaluation, some user captures failed due to insufficient data to align the model. Given that proper alignment and registration are essential for creating a successful SplatOverflow scene, we plan to improve feedback in SplatOverflow's mobile interface. The capture process could also be enhanced by allowing users to specify the hardware they intend to scan. With this information, the interface can guide users through capturing enough frames to align the splat to the CAD model and indicate which \textit{grounding} and \textit{constraint} tags require more footage.

\subsubsection{Automatically Preparing CAD Models}
To make a CAD model compatible with SplatOverflow, the producers of the hardware need to determine optimal locations to place the \textit{grounding}
and \textit{constraint} tags. The choice of location for these tags can affect how difficult it is to align the captured splat to the CAD model. To support onboarding hardware into SplatOverflow, we plan to automate the tag placement process to optimize for visibility using a process similar to \textit{BrightMarker} \cite{dogan_brightmarker_2023}. 
 
\subsubsection{IP of CAD models}
In some cases, CAD models are highly protected by hardware maintainers, who may not want to share these files with users. Currently, SplatOverflow uses a mesh representation of the CAD model and only requires information about the position of alignment tags on the model. For maintainers concerned about sharing mesh models, we plan to build an import pipeline that strips only the relevant data for our workflows without saving the actual CAD model on our server.

\subsubsection{Mixed Reality}
SplatOverflow is built in the browser and could be extended to fully immersive Augmented and Virtual Reality contexts by leveraging the WebXR Devices API~\cite{maclntyre2018thoughts}. While our implementation focuses on screen-based interaction, future work may examine how authoring SplatOverflow gestures could be extended to mixed reality. 

\section{Limitations}
This section details some limitations we have observed over the course of designing and implementing SplatOverflow.

\subsection{Acquiring CAD Models}
SplatOverflow requires access to the hardware's CAD model to populate the scene. For SplatOverflow scenes to be useful, the CAD models must also be \textit{complete}. CAD models like this can be challenging to obtain outside of open-source hardware projects, which can limit the adoption of SplatOverflow. Moreover, even when CAD files are available, the structure of sub-assemblies, parts, and components is not standardized. As a result, preparing CAD files for use in SplatOverflow can require coercing the structure of the CAD assembly into a compatible shape.

\subsection{Structure References in Technical Documentation}
Another limitation of SplatOverflow lies in its ability to reference a corpus of existing technical documentation. With the Lumen v3, each technical documentation section contained links referencing the related parts. This greatly simplified populating the documentation interface with instructions to search through. However, not all hardware will share technical documentation in the same structured fields as Opulo does with Lumen v3. As a result, maintainers must post-process existing technical documentation to add the relevant structured data indicating which CAD components are referenced. 

\subsection{Errors in Tag Placement}
In all our examples, \textit{grounding} and \textit{constraint} tags were manually placed on the hardware. In doing so, we realized that while getting precisely aligned CAD models using manually placed tags is possible, the process is also prone to errors. Specifically, if a tag is misplaced, the amount SplatOverflow's alignment deviates increases proportionate to the distance from the tag. As a result, placing SplatOverflow tags manually is a challenging, error-prone task and may not be approachable for novice users. 
\section{Conclusion}
We introduced SplatOverflow, a novel \change{workflow} for asynchronous hardware troubleshooting. SplatOverflow constructs a boundary object that can capture the physical details of a user's hardware and communicate instructions as actions on their hardware. We demonstrate SplatOverflow through a series of \change{examples} with different kinds of hardware. We show how these workflows can support complex troubleshooting workflows that require multiple exchanges between users and physical manipulation of the hardware. Moreover, we illustrate how a SplatOverflow scene can index technical knowledge about hardware to the CAD model. As a result, local users searching multiple disparate sources for relevant information can now access documentation and past solutions directly through the physical hardware.

We plan to distribute and deploy SplatOverflow with a broader user base through hardware maintainers and study our proposed forms of collaboration \textit{in the wild.} Besides direct utility for hardware users and maintainers, we believe tools for communicating about physical hardware can reduce barriers to entry for smaller producers distributing niche hardware and supporting community development amongst their users.

\begin{acks}
We thank the Digital Life Initiative at Cornell Tech for supporting this work through a doctoral fellowship. We thank the Bowers CIS Undergraduate Research Experience for supporting this work through their summer research program. We thank Roy Zunder for helping review this manuscript. Finally, we thank Joey Castillo, Frank Bu, and Stephen Hawes for participating in preliminary discussions that helped motivate this work.
\end{acks}

\bibliographystyle{ACM-Reference-Format}
\bibliography{references}


\begin{thebibliography}{79}

%%% the Web address.

\ifx \showCODEN    \undefined \def \showCODEN     #1{\unskip}     \fi
\ifx \showDOI      \undefined \def \showDOI       #1{#1}\fi
\ifx \showISBNx    \undefined \def \showISBNx     #1{\unskip}     \fi
\ifx \showISBNxiii \undefined \def \showISBNxiii  #1{\unskip}     \fi
\ifx \showISSN     \undefined \def \showISSN      #1{\unskip}     \fi
\ifx \showLCCN     \undefined \def \showLCCN      #1{\unskip}     \fi
\ifx \shownote     \undefined \def \shownote      #1{#1}          \fi
\ifx \showarticletitle \undefined \def \showarticletitle #1{#1}   \fi
\ifx \showURL      \undefined \def \showURL       {\relax}        \fi
% invisible to TeX
\providecommand\bibfield[2]{#2}
\providecommand\bibinfo[2]{#2}
\providecommand\natexlab[1]{#1}
\providecommand\showeprint[2][]{arXiv:#2}

\bibitem[3D(2023)]%
        {beltTension}
\bibfield{author}{\bibinfo{person}{Prusa 3D}.} \bibinfo{year}{2023}\natexlab{}.
\newblock \bibinfo{title}{Adjusting the belt tension on the Original Prusa MK4 - Belt Tuner App}.
\newblock
\newblock
\urldef\tempurl%
\url{https://youtu.be/oeq2MVxE_H8?feature=shared&t=56}
\showURL{%
\tempurl}


\bibitem[Ackerman et~al\mbox{.}(2013)]%
        {ackerman_sharing_2013}
\bibfield{author}{\bibinfo{person}{Mark~S. Ackerman}, \bibinfo{person}{Juri Dachtera}, \bibinfo{person}{Volkmar Pipek}, {and} \bibinfo{person}{Volker Wulf}.} \bibinfo{year}{2013}\natexlab{}.
\newblock \showarticletitle{Sharing {Knowledge} and {Expertise}: {The} {CSCW} {View} of {Knowledge} {Management}}.
\newblock \bibinfo{journal}{\emph{Computer Supported Cooperative Work (CSCW)}} \bibinfo{volume}{22}, \bibinfo{number}{4-6} (\bibinfo{date}{Aug.} \bibinfo{year}{2013}), \bibinfo{pages}{531--573}.
\newblock
\showISSN{0925-9724, 1573-7551}
\urldef\tempurl%
\url{https://doi.org/10.1007/s10606-013-9192-8}
\showDOI{\tempurl}


\bibitem[Ackermann(2008)]%
        {ackermann_toward_2008}
\bibfield{author}{\bibinfo{person}{John~R Ackermann}.} \bibinfo{year}{2008}\natexlab{}.
\newblock \showarticletitle{Toward {Open} {Source} {Hardware}}.
\newblock  (\bibinfo{year}{2008}).
\newblock


\bibitem[Agrawala et~al\mbox{.}(2003)]%
        {agrawala-assembly-instructions}
\bibfield{author}{\bibinfo{person}{Maneesh Agrawala}, \bibinfo{person}{Doantam Phan}, \bibinfo{person}{Julie Heiser}, \bibinfo{person}{John Haymaker}, \bibinfo{person}{Jeff Klingner}, \bibinfo{person}{Pat Hanrahan}, {and} \bibinfo{person}{Barbara Tversky}.} \bibinfo{year}{2003}\natexlab{}.
\newblock \showarticletitle{Designing Effective Step-by-Step Assembly Instructions}.
\newblock \bibinfo{journal}{\emph{ACM Trans. Graph.}} \bibinfo{volume}{22}, \bibinfo{number}{3} (\bibinfo{date}{jul} \bibinfo{year}{2003}), \bibinfo{pages}{828–837}.
\newblock
\showISSN{0730-0301}
\urldef\tempurl%
\url{https://doi.org/10.1145/882262.882352}
\showDOI{\tempurl}


\bibitem[Anderson et~al\mbox{.}(2012)]%
        {anderson_discovering_2012}
\bibfield{author}{\bibinfo{person}{Ashton Anderson}, \bibinfo{person}{Daniel Huttenlocher}, \bibinfo{person}{Jon Kleinberg}, {and} \bibinfo{person}{Jure Leskovec}.} \bibinfo{year}{2012}\natexlab{}.
\newblock \showarticletitle{Discovering value from community activity on focused question answering sites: a case study of stack overflow}. In \bibinfo{booktitle}{\emph{Proceedings of the 18th {ACM} {SIGKDD} international conference on {Knowledge} discovery and data mining}}. \bibinfo{publisher}{ACM}, \bibinfo{address}{Beijing China}, \bibinfo{pages}{850--858}.
\newblock
\showISBNx{978-1-4503-1462-6}
\urldef\tempurl%
\url{https://doi.org/10.1145/2339530.2339665}
\showDOI{\tempurl}


\bibitem[Arun et~al\mbox{.}(1987)]%
        {arun_least-squares_1987}
\bibfield{author}{\bibinfo{person}{K.~S. Arun}, \bibinfo{person}{T.~S. Huang}, {and} \bibinfo{person}{S.~D. Blostein}.} \bibinfo{year}{1987}\natexlab{}.
\newblock \showarticletitle{Least-{Squares} {Fitting} of {Two} 3-{D} {Point} {Sets}}.
\newblock \bibinfo{journal}{\emph{IEEE Transactions on Pattern Analysis and Machine Intelligence}} \bibinfo{volume}{PAMI-9}, \bibinfo{number}{5} (\bibinfo{date}{Sept.} \bibinfo{year}{1987}), \bibinfo{pages}{698--700}.
\newblock
\showISSN{0162-8828}
\urldef\tempurl%
\url{https://doi.org/10.1109/TPAMI.1987.4767965}
\showDOI{\tempurl}


\bibitem[Benkler(2002)]%
        {benkler2002coase}
\bibfield{author}{\bibinfo{person}{Yochai Benkler}.} \bibinfo{year}{2002}\natexlab{}.
\newblock \showarticletitle{Coase's penguin, or, linux and" the nature of the firm"}.
\newblock \bibinfo{journal}{\emph{Yale law journal}} (\bibinfo{year}{2002}), \bibinfo{pages}{369--446}.
\newblock


\bibitem[Benkler and Nissenbaum(2006)]%
        {benkler2006commons}
\bibfield{author}{\bibinfo{person}{Yochai Benkler} {and} \bibinfo{person}{Helen Nissenbaum}.} \bibinfo{year}{2006}\natexlab{}.
\newblock \showarticletitle{Commons-based peer production and virtue}.
\newblock \bibinfo{journal}{\emph{Journal of political philosophy}} \bibinfo{volume}{14}, \bibinfo{number}{4} (\bibinfo{year}{2006}).
\newblock


\bibitem[Bhatia(2018)]%
        {bhatia2018interdiscursive}
\bibfield{author}{\bibinfo{person}{Aditi Bhatia}.} \bibinfo{year}{2018}\natexlab{}.
\newblock \showarticletitle{Interdiscursive performance in digital professions: The case of YouTube tutorials}.
\newblock \bibinfo{journal}{\emph{Journal of Pragmatics}}  \bibinfo{volume}{124} (\bibinfo{year}{2018}), \bibinfo{pages}{106--120}.
\newblock


\bibitem[Bobrow and Whalen(2002)]%
        {bobrow_community_2002}
\bibfield{author}{\bibinfo{person}{Daniel~G. Bobrow} {and} \bibinfo{person}{Jack Whalen}.} \bibinfo{year}{2002}\natexlab{}.
\newblock \showarticletitle{Community {Knowledge} {Sharing} in {Practice}: {The} {Eureka} {Story}}.
\newblock \bibinfo{journal}{\emph{Reflections: The SoL Journal}} \bibinfo{volume}{4}, \bibinfo{number}{2} (\bibinfo{date}{Dec.} \bibinfo{year}{2002}), \bibinfo{pages}{47--59}.
\newblock
\showISSN{15360148, 15241734}
\urldef\tempurl%
\url{https://doi.org/10.1162/152417302762251336}
\showDOI{\tempurl}


\bibitem[Bonvoisin et~al\mbox{.}(2017)]%
        {bonvoisin_what_2017}
\bibfield{author}{\bibinfo{person}{Jérémy Bonvoisin}, \bibinfo{person}{Robert Mies}, \bibinfo{person}{Jean-François Boujut}, {and} \bibinfo{person}{Rainer Stark}.} \bibinfo{year}{2017}\natexlab{}.
\newblock \showarticletitle{What is the “{Source}” of {Open} {Source} {Hardware}?}
\newblock \bibinfo{journal}{\emph{Journal of Open Hardware}} \bibinfo{volume}{1}, \bibinfo{number}{1} (\bibinfo{date}{Sept.} \bibinfo{year}{2017}), \bibinfo{pages}{5}.
\newblock
\showISSN{2514-1708}
\urldef\tempurl%
\url{https://doi.org/10.5334/joh.7}
\showDOI{\tempurl}


\bibitem[Brand(2025)]%
        {Brand2025-hc}
\bibfield{author}{\bibinfo{person}{Stewart Brand}.} \bibinfo{year}{2025}\natexlab{}.
\newblock \bibinfo{booktitle}{\emph{Maintenance: {Of} everything: {Part} one}}.
\newblock \bibinfo{publisher}{Stripe Press}.
\newblock


\bibitem[Brooke({[n.\,d.]})]%
        {brooke_sus_nodate}
\bibfield{author}{\bibinfo{person}{John Brooke}.} \bibinfo{year}{[n.\,d.]}\natexlab{}.
\newblock \showarticletitle{{SUS} - {A} quick and dirty usability scale}.
\newblock  (\bibinfo{year}{[n.\,d.]}).
\newblock


\bibitem[Buxton(1992)]%
        {buxton1992telepresence}
\bibfield{author}{\bibinfo{person}{William Buxton}.} \bibinfo{year}{1992}\natexlab{}.
\newblock \showarticletitle{Telepresence: Integrating shared task and person spaces}. In \bibinfo{booktitle}{\emph{Proceedings of graphics interface}}, Vol.~\bibinfo{volume}{92}. Canadian Information Processing Society Toronto, Canada, \bibinfo{pages}{123--129}.
\newblock


\bibitem[Castillo({[n.\,d.]})]%
        {castillo_open_nodate}
\bibfield{author}{\bibinfo{person}{Joey Castillo}.} \bibinfo{year}{[n.\,d.]}\natexlab{}.
\newblock \bibinfo{title}{Open {Book} {Project}}.
\newblock
\newblock
\urldef\tempurl%
\url{https://github.com/joeycastillo/The-Open-Book/tree/reboot#state-of-the-book}
\showURL{%
\tempurl}


\bibitem[Chastine et~al\mbox{.}(2006)]%
        {chastine-inter-referential}
\bibfield{author}{\bibinfo{person}{Jeffrey~W. Chastine}, \bibinfo{person}{Ying Zhu}, {and} \bibinfo{person}{Jon~A. Preston}.} \bibinfo{year}{2006}\natexlab{}.
\newblock \showarticletitle{A Framework for Inter-referential Awareness in Collaborative Environments}. In \bibinfo{booktitle}{\emph{2006 International Conference on Collaborative Computing: Networking, Applications and Worksharing}}. \bibinfo{pages}{1--5}.
\newblock
\urldef\tempurl%
\url{https://doi.org/10.1109/COLCOM.2006.361859}
\showDOI{\tempurl}


\bibitem[Chatterjee et~al\mbox{.}(2022)]%
        {chatterjee_ardw_2022}
\bibfield{author}{\bibinfo{person}{Ishan Chatterjee}, \bibinfo{person}{Tadeusz Pforte}, \bibinfo{person}{Aspen Tng}, \bibinfo{person}{Farshid Salemi~Parizi}, \bibinfo{person}{Chaoran Chen}, {and} \bibinfo{person}{Shwetak Patel}.} \bibinfo{year}{2022}\natexlab{}.
\newblock \showarticletitle{{ARDW}: {An} {Augmented} {Reality} {Workbench} for {Printed} {Circuit} {Board} {Debugging}}. In \bibinfo{booktitle}{\emph{Proceedings of the 35th {Annual} {ACM} {Symposium} on {User} {Interface} {Software} and {Technology}}}. \bibinfo{publisher}{ACM}, \bibinfo{address}{Bend OR USA}, \bibinfo{pages}{1--16}.
\newblock
\showISBNx{978-1-4503-9320-1}
\urldef\tempurl%
\url{https://doi.org/10.1145/3526113.3545684}
\showDOI{\tempurl}


\bibitem[Chen et~al\mbox{.}(2021)]%
        {collab-pointing-ar}
\bibfield{author}{\bibinfo{person}{Lei Chen}, \bibinfo{person}{Yilin Liu}, \bibinfo{person}{Yue Li}, \bibinfo{person}{Lingyun Yu}, \bibinfo{person}{BoYu Gao}, \bibinfo{person}{Maurizio Caon}, \bibinfo{person}{Yong Yue}, {and} \bibinfo{person}{Hai-Ning Liang}.} \bibinfo{year}{2021}\natexlab{}.
\newblock \showarticletitle{Effect of Visual Cues on Pointing Tasks in Co-Located Augmented Reality Collaboration}. In \bibinfo{booktitle}{\emph{Proceedings of the 2021 ACM Symposium on Spatial User Interaction}} (Virtual Event, USA) \emph{(\bibinfo{series}{SUI '21})}. \bibinfo{publisher}{Association for Computing Machinery}, \bibinfo{address}{New York, NY, USA}, Article \bibinfo{articleno}{12}, \bibinfo{numpages}{12}~pages.
\newblock
\showISBNx{9781450390910}
\urldef\tempurl%
\url{https://doi.org/10.1145/3485279.3485297}
\showDOI{\tempurl}


\bibitem[Dabbish et~al\mbox{.}(2012)]%
        {dabbish2012social}
\bibfield{author}{\bibinfo{person}{Laura Dabbish}, \bibinfo{person}{Colleen Stuart}, \bibinfo{person}{Jason Tsay}, {and} \bibinfo{person}{Jim Herbsleb}.} \bibinfo{year}{2012}\natexlab{}.
\newblock \showarticletitle{Social coding in GitHub: transparency and collaboration in an open software repository}. In \bibinfo{booktitle}{\emph{Proceedings of the ACM 2012 conference on computer supported cooperative work}}. \bibinfo{pages}{1277--1286}.
\newblock


\bibitem[Danchilla(2012)]%
        {danchilla_threejs_2012}
\bibfield{author}{\bibinfo{person}{Brian Danchilla}.} \bibinfo{year}{2012}\natexlab{}.
\newblock \showarticletitle{Three.js {Framework}}.
\newblock In \bibinfo{booktitle}{\emph{Beginning {WebGL} for {HTML5}}}. \bibinfo{publisher}{Apress}, \bibinfo{address}{Berkeley, CA}, \bibinfo{pages}{173--203}.
\newblock
\showISBNx{978-1-4302-3996-3 978-1-4302-3997-0}
\urldef\tempurl%
\url{https://doi.org/10.1007/978-1-4302-3997-0_7}
\showDOI{\tempurl}


\bibitem[Dogan et~al\mbox{.}(2023a)]%
        {dogan_structcode_2023}
\bibfield{author}{\bibinfo{person}{Mustafa~Doga Dogan}, \bibinfo{person}{Vivian~Hsinyueh Chan}, \bibinfo{person}{Richard Qi}, \bibinfo{person}{Grace Tang}, \bibinfo{person}{Thijs Roumen}, {and} \bibinfo{person}{Stefanie Mueller}.} \bibinfo{year}{2023}\natexlab{a}.
\newblock \showarticletitle{{StructCode}: {Leveraging} {Fabrication} {Artifacts} to {Store} {Data} in {Laser}-{Cut} {Objects}}. In \bibinfo{booktitle}{\emph{Proceedings of the 8th {ACM} {Symposium} on {Computational} {Fabrication}}} \emph{(\bibinfo{series}{{SCF} '23})}. \bibinfo{publisher}{Association for Computing Machinery}, \bibinfo{address}{New York, NY, USA}, \bibinfo{pages}{1--13}.
\newblock
\showISBNx{9798400703195}
\urldef\tempurl%
\url{https://doi.org/10.1145/3623263.3623353}
\showDOI{\tempurl}


\bibitem[Dogan et~al\mbox{.}(2023b)]%
        {dogan_brightmarker_2023}
\bibfield{author}{\bibinfo{person}{Mustafa~Doga Dogan}, \bibinfo{person}{Raul Garcia-Martin}, \bibinfo{person}{Patrick~William Haertel}, \bibinfo{person}{Jamison~John O'Keefe}, \bibinfo{person}{Ahmad Taka}, \bibinfo{person}{Akarsh Aurora}, \bibinfo{person}{Raul Sanchez-Reillo}, {and} \bibinfo{person}{Stefanie Mueller}.} \bibinfo{year}{2023}\natexlab{b}.
\newblock \showarticletitle{{BrightMarker}: {3D} {Printed} {Fluorescent} {Markers} for {Object} {Tracking}}. In \bibinfo{booktitle}{\emph{Proceedings of the 36th {Annual} {ACM} {Symposium} on {User} {Interface} {Software} and {Technology}}}. \bibinfo{publisher}{ACM}, \bibinfo{address}{San Francisco CA USA}, \bibinfo{pages}{1--13}.
\newblock
\showISBNx{9798400701320}
\urldef\tempurl%
\url{https://doi.org/10.1145/3586183.3606758}
\showDOI{\tempurl}


\bibitem[Dogan et~al\mbox{.}(2022)]%
        {dogan_infraredtags_2022}
\bibfield{author}{\bibinfo{person}{Mustafa~Doga Dogan}, \bibinfo{person}{Ahmad Taka}, \bibinfo{person}{Michael Lu}, \bibinfo{person}{Yunyi Zhu}, \bibinfo{person}{Akshat Kumar}, \bibinfo{person}{Aakar Gupta}, {and} \bibinfo{person}{Stefanie Mueller}.} \bibinfo{year}{2022}\natexlab{}.
\newblock \showarticletitle{{InfraredTags}: {Embedding} {Invisible} {AR} {Markers} and {Barcodes} {Using} {Low}-{Cost}, {Infrared}-{Based} {3D} {Printing} and {Imaging} {Tools}}. In \bibinfo{booktitle}{\emph{Proceedings of the 2022 {CHI} {Conference} on {Human} {Factors} in {Computing} {Systems}}} \emph{(\bibinfo{series}{{CHI} '22})}. \bibinfo{publisher}{Association for Computing Machinery}, \bibinfo{address}{New York, NY, USA}.
\newblock
\showISBNx{978-1-4503-9157-3}
\urldef\tempurl%
\url{https://doi.org/10.1145/3491102.3501951}
\showDOI{\tempurl}
\newblock
\shownote{event-place: New Orleans, LA, USA}.


\bibitem[Dourish and Bellotti(1992)]%
        {dourish_awareness_1992}
\bibfield{author}{\bibinfo{person}{Paul Dourish} {and} \bibinfo{person}{Victoria Bellotti}.} \bibinfo{year}{1992}\natexlab{}.
\newblock \showarticletitle{Awareness and coordination in shared workspaces}. In \bibinfo{booktitle}{\emph{Proceedings of the 1992 {ACM} conference on {Computer}-supported cooperative work - {CSCW} '92}}. \bibinfo{publisher}{ACM Press}, \bibinfo{address}{Toronto, Ontario, Canada}, \bibinfo{pages}{107--114}.
\newblock
\showISBNx{978-0-89791-542-7}
\urldef\tempurl%
\url{https://doi.org/10.1145/143457.143468}
\showDOI{\tempurl}


\bibitem[Dunn et~al\mbox{.}(2023)]%
        {dunn_jubilee_2023}
\bibfield{author}{\bibinfo{person}{Kellie Dunn}, \bibinfo{person}{Cynthia Feng}, {and} \bibinfo{person}{Nadya Peek}.} \bibinfo{year}{2023}\natexlab{}.
\newblock \showarticletitle{Jubilee: {A} {Case} {Study} of {Distributed} {Manufacturing} in an {Open} {Source} {Hardware} {Project}}.
\newblock \bibinfo{journal}{\emph{Journal of Open Hardware}} \bibinfo{volume}{7}, \bibinfo{number}{1} (\bibinfo{date}{May} \bibinfo{year}{2023}), \bibinfo{pages}{4}.
\newblock
\showISSN{2514-1708}
\urldef\tempurl%
\url{https://doi.org/10.5334/joh.51}
\showDOI{\tempurl}


\bibitem[Eghbal(2020)]%
        {eghbal2020working}
\bibfield{author}{\bibinfo{person}{Nadia Eghbal}.} \bibinfo{year}{2020}\natexlab{}.
\newblock \bibinfo{booktitle}{\emph{Working in public: the making and maintenance of open source software}}.
\newblock \bibinfo{publisher}{Stripe Press}.
\newblock


\bibitem[Fussell et~al\mbox{.}(2000a)]%
        {fussell-coordination}
\bibfield{author}{\bibinfo{person}{Susan~R. Fussell}, \bibinfo{person}{Robert~E. Kraut}, {and} \bibinfo{person}{Jane Siegel}.} \bibinfo{year}{2000}\natexlab{a}.
\newblock \showarticletitle{Coordination of Communication: Effects of Shared Visual Context on Collaborative Work}. In \bibinfo{booktitle}{\emph{Proceedings of the 2000 ACM Conference on Computer Supported Cooperative Work}} (Philadelphia, Pennsylvania, USA) \emph{(\bibinfo{series}{CSCW '00})}. \bibinfo{publisher}{Association for Computing Machinery}, \bibinfo{address}{New York, NY, USA}, \bibinfo{pages}{21–30}.
\newblock
\showISBNx{1581132220}
\urldef\tempurl%
\url{https://doi.org/10.1145/358916.358947}
\showDOI{\tempurl}


\bibitem[Fussell et~al\mbox{.}(2000b)]%
        {fussell_coordination_2000}
\bibfield{author}{\bibinfo{person}{Susan~R. Fussell}, \bibinfo{person}{Robert~E. Kraut}, {and} \bibinfo{person}{Jane Siegel}.} \bibinfo{year}{2000}\natexlab{b}.
\newblock \showarticletitle{Coordination of communication: effects of shared visual context on collaborative work}. In \bibinfo{booktitle}{\emph{Proceedings of the 2000 {ACM} conference on {Computer} supported cooperative work}}. \bibinfo{publisher}{ACM}, \bibinfo{address}{Philadelphia Pennsylvania USA}, \bibinfo{pages}{21--30}.
\newblock
\showISBNx{978-1-58113-222-9}
\urldef\tempurl%
\url{https://doi.org/10.1145/358916.358947}
\showDOI{\tempurl}


\bibitem[Gauglitz et~al\mbox{.}(2012)]%
        {gauglitz_integrating_2012}
\bibfield{author}{\bibinfo{person}{Steffen Gauglitz}, \bibinfo{person}{Cha Lee}, \bibinfo{person}{Matthew Turk}, {and} \bibinfo{person}{Tobias Höllerer}.} \bibinfo{year}{2012}\natexlab{}.
\newblock \showarticletitle{Integrating the {Physical} {Environment} into {Mobile} {Remote} {Collaboration}}. In \bibinfo{booktitle}{\emph{Proceedings of the 14th {International} {Conference} on {Human}-{Computer} {Interaction} with {Mobile} {Devices} and {Services}}} \emph{(\bibinfo{series}{{MobileHCI} '12})}. \bibinfo{publisher}{Association for Computing Machinery}, \bibinfo{address}{New York, NY, USA}, \bibinfo{pages}{241--250}.
\newblock
\showISBNx{978-1-4503-1105-2}
\urldef\tempurl%
\url{https://doi.org/10.1145/2371574.2371610}
\showDOI{\tempurl}
\newblock
\shownote{event-place: San Francisco, California, USA}.


\bibitem[Gauglitz et~al\mbox{.}(2014)]%
        {gauglitz_world-stabilized_2014}
\bibfield{author}{\bibinfo{person}{Steffen Gauglitz}, \bibinfo{person}{Benjamin Nuernberger}, \bibinfo{person}{Matthew Turk}, {and} \bibinfo{person}{Tobias Höllerer}.} \bibinfo{year}{2014}\natexlab{}.
\newblock \showarticletitle{World-stabilized annotations and virtual scene navigation for remote collaboration}. In \bibinfo{booktitle}{\emph{Proceedings of the 27th annual {ACM} symposium on {User} interface software and technology}}. \bibinfo{publisher}{ACM}, \bibinfo{address}{Honolulu Hawaii USA}, \bibinfo{pages}{449--459}.
\newblock
\showISBNx{978-1-4503-3069-5}
\urldef\tempurl%
\url{https://doi.org/10.1145/2642918.2647372}
\showDOI{\tempurl}


\bibitem[Gibb(2015)]%
        {gibb2015building}
\bibfield{author}{\bibinfo{person}{Alicia Gibb}.} \bibinfo{year}{2015}\natexlab{}.
\newblock \bibinfo{booktitle}{\emph{Building open source hardware: DIY manufacturing for hackers and makers}}.
\newblock \bibinfo{publisher}{Pearson Education}.
\newblock


\bibitem[Gutwin and Greenberg(2002)]%
        {gutwin_descriptive_2002}
\bibfield{author}{\bibinfo{person}{Carl Gutwin} {and} \bibinfo{person}{Saul Greenberg}.} \bibinfo{year}{2002}\natexlab{}.
\newblock \showarticletitle{A {Descriptive} {Framework} of {Workspace} {Awareness} for {Real}-{Time} {Groupware}}.
\newblock \bibinfo{journal}{\emph{Computer Supported Cooperative Work (CSCW)}} \bibinfo{volume}{11}, \bibinfo{number}{3-4} (\bibinfo{date}{Sept.} \bibinfo{year}{2002}), \bibinfo{pages}{411--446}.
\newblock
\showISSN{0925-9724, 1573-7551}
\urldef\tempurl%
\url{https://doi.org/10.1023/A:1021271517844}
\showDOI{\tempurl}


\bibitem[Guédon and Lepetit(2023)]%
        {guedon_sugar_2023}
\bibfield{author}{\bibinfo{person}{Antoine Guédon} {and} \bibinfo{person}{Vincent Lepetit}.} \bibinfo{year}{2023}\natexlab{}.
\newblock \showarticletitle{{SuGaR}: {Surface}-{Aligned} {Gaussian} {Splatting} for {Efficient} {3D} {Mesh} {Reconstruction} and {High}-{Quality} {Mesh} {Rendering}}.
\newblock  (\bibinfo{year}{2023}).
\newblock
\urldef\tempurl%
\url{https://doi.org/10.48550/ARXIV.2311.12775}
\showDOI{\tempurl}
\newblock
\shownote{Publisher: [object Object] Version Number: 3}.


\bibitem[Hackaday(2018)]%
        {htmlBOM}
\bibfield{author}{\bibinfo{person}{Hackaday}.} \bibinfo{year}{2018}\natexlab{}.
\newblock \bibinfo{title}{Interactive KiCAD BOMs Make Hand Assembly A Breeze}.
\newblock
\newblock
\urldef\tempurl%
\url{https://hackaday.com/2018/09/04/interactive-kicad-boms-make-hand-assembly-a-breeze/}
\showURL{%
\tempurl}


\bibitem[Hodges and Chen(2019)]%
        {hodges_long_2019}
\bibfield{author}{\bibinfo{person}{Steve Hodges} {and} \bibinfo{person}{Nicholas Chen}.} \bibinfo{year}{2019}\natexlab{}.
\newblock \showarticletitle{Long {Tail} {Hardware}: {Turning} {Device} {Concepts} {Into} {Viable} {Low} {Volume} {Products}}.
\newblock \bibinfo{journal}{\emph{IEEE Pervasive Computing}} \bibinfo{volume}{18}, \bibinfo{number}{4} (\bibinfo{date}{Oct.} \bibinfo{year}{2019}), \bibinfo{pages}{51--59}.
\newblock
\showISSN{1558-2590}
\urldef\tempurl%
\url{https://doi.org/10.1109/MPRV.2019.2947966}
\showDOI{\tempurl}
\newblock
\shownote{Conference Name: IEEE Pervasive Computing}.


\bibitem[Izadi et~al\mbox{.}(2011)]%
        {izadi_kinectfusion_2011}
\bibfield{author}{\bibinfo{person}{Shahram Izadi}, \bibinfo{person}{David Kim}, \bibinfo{person}{Otmar Hilliges}, \bibinfo{person}{David Molyneaux}, \bibinfo{person}{Richard Newcombe}, \bibinfo{person}{Pushmeet Kohli}, \bibinfo{person}{Jamie Shotton}, \bibinfo{person}{Steve Hodges}, \bibinfo{person}{Dustin Freeman}, \bibinfo{person}{Andrew Davison}, {and} \bibinfo{person}{Andrew Fitzgibbon}.} \bibinfo{year}{2011}\natexlab{}.
\newblock \showarticletitle{{KinectFusion}: real-time {3D} reconstruction and interaction using a moving depth camera}. In \bibinfo{booktitle}{\emph{Proceedings of the 24th annual {ACM} symposium on {User} interface software and technology}}. \bibinfo{publisher}{ACM}, \bibinfo{address}{Santa Barbara California USA}, \bibinfo{pages}{559--568}.
\newblock
\showISBNx{978-1-4503-0716-1}
\urldef\tempurl%
\url{https://doi.org/10.1145/2047196.2047270}
\showDOI{\tempurl}


\bibitem[Jackson(2014)]%
        {jackson_rethinking_2014}
\bibfield{author}{\bibinfo{person}{Steven~J. Jackson}.} \bibinfo{year}{2014}\natexlab{}.
\newblock \showarticletitle{Rethinking {Repair}}.
\newblock In \bibinfo{booktitle}{\emph{Media {Technologies}: {Essays} on {Communication}, {Materiality}, and {Society}}}, \bibfield{editor}{\bibinfo{person}{Tarleton Gillespie}, \bibinfo{person}{Pablo~J. Boczkowski}, {and} \bibinfo{person}{Kirsten~A. Foot}} (Eds.). \bibinfo{publisher}{The MIT Press}, \bibinfo{pages}{0}.
\newblock
\showISBNx{978-0-262-52537-4}
\urldef\tempurl%
\url{https://doi.org/10.7551/mitpress/9780262525374.003.0011}
\showDOI{\tempurl}


\bibitem[Johnson et~al\mbox{.}(2021)]%
        {johnson-know-where-that-is}
\bibfield{author}{\bibinfo{person}{Janet~G Johnson}, \bibinfo{person}{Danilo Gasques}, \bibinfo{person}{Tommy Sharkey}, \bibinfo{person}{Evan Schmitz}, {and} \bibinfo{person}{Nadir Weibel}.} \bibinfo{year}{2021}\natexlab{}.
\newblock \showarticletitle{Do You Really Need to Know Where “That” Is? Enhancing Support for Referencing in Collaborative Mixed Reality Environments}. In \bibinfo{booktitle}{\emph{Proceedings of the 2021 CHI Conference on Human Factors in Computing Systems}} (Yokohama, Japan) \emph{(\bibinfo{series}{CHI '21})}. \bibinfo{publisher}{Association for Computing Machinery}, \bibinfo{address}{New York, NY, USA}, Article \bibinfo{articleno}{514}, \bibinfo{numpages}{14}~pages.
\newblock
\showISBNx{9781450380966}
\urldef\tempurl%
\url{https://doi.org/10.1145/3411764.3445246}
\showDOI{\tempurl}


\bibitem[Karchemsky et~al\mbox{.}(2019)]%
        {Heimdall}
\bibfield{author}{\bibinfo{person}{Mitchell Karchemsky}, \bibinfo{person}{J.D. Zamfirescu-Pereira}, \bibinfo{person}{Kuan-Ju Wu}, \bibinfo{person}{Fran\c{c}ois Guimbreti\`{e}re}, {and} \bibinfo{person}{Bjoern Hartmann}.} \bibinfo{year}{2019}\natexlab{}.
\newblock \showarticletitle{Heimdall: A Remotely Controlled Inspection Workbench For Debugging Microcontroller Projects}. In \bibinfo{booktitle}{\emph{Proceedings of the 2019 CHI Conference on Human Factors in Computing Systems}} (Glasgow, Scotland Uk) \emph{(\bibinfo{series}{CHI '19})}. \bibinfo{publisher}{Association for Computing Machinery}, \bibinfo{address}{New York, NY, USA}, \bibinfo{pages}{1–12}.
\newblock
\showISBNx{9781450359702}
\urldef\tempurl%
\url{https://doi.org/10.1145/3290605.3300728}
\showDOI{\tempurl}


\bibitem[Kellog(2024)]%
        {kellog_3d_2024}
\bibfield{author}{\bibinfo{person}{Mark Kellog}.} \bibinfo{year}{2024}\natexlab{}.
\newblock \bibinfo{title}{{3D} {Gaussian} splatting for {Three}.js}.
\newblock
\newblock
\urldef\tempurl%
\url{https://github.com/mkkellogg/GaussianSplats3D}
\showURL{%
\tempurl}


\bibitem[Kerbl et~al\mbox{.}(2023)]%
        {kerbl_3d_2023}
\bibfield{author}{\bibinfo{person}{Bernhard Kerbl}, \bibinfo{person}{Georgios Kopanas}, \bibinfo{person}{Thomas Leimkuehler}, {and} \bibinfo{person}{George Drettakis}.} \bibinfo{year}{2023}\natexlab{}.
\newblock \showarticletitle{{3D} {Gaussian} {Splatting} for {Real}-{Time} {Radiance} {Field} {Rendering}}.
\newblock \bibinfo{journal}{\emph{ACM Transactions on Graphics}} \bibinfo{volume}{42}, \bibinfo{number}{4} (\bibinfo{date}{Aug.} \bibinfo{year}{2023}), \bibinfo{pages}{1--14}.
\newblock
\showISSN{0730-0301, 1557-7368}
\urldef\tempurl%
\url{https://doi.org/10.1145/3592433}
\showDOI{\tempurl}


\bibitem[Kim et~al\mbox{.}(2014)]%
        {kim2014crowdsourcing}
\bibfield{author}{\bibinfo{person}{Juho Kim}, \bibinfo{person}{Phu~Tran Nguyen}, \bibinfo{person}{Sarah Weir}, \bibinfo{person}{Philip~J Guo}, \bibinfo{person}{Robert~C Miller}, {and} \bibinfo{person}{Krzysztof~Z Gajos}.} \bibinfo{year}{2014}\natexlab{}.
\newblock \showarticletitle{Crowdsourcing step-by-step information extraction to enhance existing how-to videos}. In \bibinfo{booktitle}{\emph{Proceedings of the SIGCHI conference on human factors in computing systems}}. \bibinfo{pages}{4017--4026}.
\newblock


\bibitem[Li et~al\mbox{.}(2023)]%
        {li_magic_2023}
\bibfield{author}{\bibinfo{person}{Ke Li}, \bibinfo{person}{Susanne Schmidt}, \bibinfo{person}{Tim Rolff}, \bibinfo{person}{Reinhard Bacher}, \bibinfo{person}{Wim Leemans}, {and} \bibinfo{person}{Frank Steinicke}.} \bibinfo{year}{2023}\natexlab{}.
\newblock \bibinfo{title}{Magic {NeRF} {Lens}: {Interactive} {Fusion} of {Neural} {Radiance} {Fields} for {Virtual} {Facility} {Inspection}}.
\newblock
\newblock
\urldef\tempurl%
\url{https://doi.org/10.48550/arXiv.2307.09860}
\showDOI{\tempurl}
\newblock
\shownote{arXiv:2307.09860 [cs]}.


\bibitem[Liu et~al\mbox{.}(2024)]%
        {liu_mvsgaussian_2024}
\bibfield{author}{\bibinfo{person}{Tianqi Liu}, \bibinfo{person}{Guangcong Wang}, \bibinfo{person}{Shoukang Hu}, \bibinfo{person}{Liao Shen}, \bibinfo{person}{Xinyi Ye}, \bibinfo{person}{Yuhang Zang}, \bibinfo{person}{Zhiguo Cao}, \bibinfo{person}{Wei Li}, {and} \bibinfo{person}{Ziwei Liu}.} \bibinfo{year}{2024}\natexlab{}.
\newblock \bibinfo{title}{{MVSGaussian}: {Fast} {Generalizable} {Gaussian} {Splatting} {Reconstruction} from {Multi}-{View} {Stereo}}.
\newblock
\newblock
\urldef\tempurl%
\url{http://arxiv.org/abs/2405.12218}
\showURL{%
\tempurl}
\newblock
\shownote{arXiv:2405.12218}.


\bibitem[Luthra and Mittal(2010)]%
        {luthra2010firebug}
\bibfield{author}{\bibinfo{person}{Chandan Luthra} {and} \bibinfo{person}{Deepak Mittal}.} \bibinfo{year}{2010}\natexlab{}.
\newblock \bibinfo{booktitle}{\emph{Firebug 1.5: Editing, Debugging, and Monitoring Web Pages}}.
\newblock \bibinfo{publisher}{Packt Publishing}.
\newblock


\bibitem[Lutters and Ackerman(2007)]%
        {lutters_beyond_2007}
\bibfield{author}{\bibinfo{person}{Wayne~G. Lutters} {and} \bibinfo{person}{Mark~S. Ackerman}.} \bibinfo{year}{2007}\natexlab{}.
\newblock \showarticletitle{Beyond {Boundary} {Objects}: {Collaborative} {Reuse} in {Aircraft} {Technical} {Support}}.
\newblock \bibinfo{journal}{\emph{Computer Supported Cooperative Work (CSCW)}} \bibinfo{volume}{16}, \bibinfo{number}{3} (\bibinfo{date}{June} \bibinfo{year}{2007}), \bibinfo{pages}{341--372}.
\newblock
\showISSN{0925-9724, 1573-7551}
\urldef\tempurl%
\url{https://doi.org/10.1007/s10606-006-9036-x}
\showDOI{\tempurl}


\bibitem[Maclntyre and Smith(2018)]%
        {maclntyre2018thoughts}
\bibfield{author}{\bibinfo{person}{Blair Maclntyre} {and} \bibinfo{person}{Trevor~F Smith}.} \bibinfo{year}{2018}\natexlab{}.
\newblock \showarticletitle{Thoughts on the Future of WebXR and the Immersive Web}. In \bibinfo{booktitle}{\emph{2018 IEEE international symposium on mixed and augmented reality adjunct (ISMAR-Adjunct)}}. IEEE, \bibinfo{pages}{338--342}.
\newblock


\bibitem[Mariscal-Melgar et~al\mbox{.}(2023)]%
        {mariscal-melgar_semi-automatic_2023}
\bibfield{author}{\bibinfo{person}{J.~C. Mariscal-Melgar}, \bibinfo{person}{Pieter Hijma}, \bibinfo{person}{Manuel Moritz}, {and} \bibinfo{person}{Tobias Redlich}.} \bibinfo{year}{2023}\natexlab{}.
\newblock \showarticletitle{Semi-{Automatic} {Generation} of {Assembly} {Instructions} for {Open} {Source} {Hardware}}.
\newblock \bibinfo{journal}{\emph{Journal of Open Hardware}} \bibinfo{volume}{7}, \bibinfo{number}{1} (\bibinfo{date}{Aug.} \bibinfo{year}{2023}), \bibinfo{pages}{6}.
\newblock
\showISSN{2514-1708}
\urldef\tempurl%
\url{https://doi.org/10.5334/joh.56}
\showDOI{\tempurl}


\bibitem[Meyer(2023)]%
        {meyer_aruco_2023}
\bibfield{author}{\bibinfo{person}{Lukas Meyer}.} \bibinfo{year}{2023}\natexlab{}.
\newblock \bibinfo{title}{Aruco {Scale} factor {Estimation} for {COLMAP}}.
\newblock
\newblock
\urldef\tempurl%
\url{https://pypi.org/project/aruco-estimator/}
\showURL{%
\tempurl}


\bibitem[Milara et~al\mbox{.}(2019)]%
        {milara_document-while-doing_2019}
\bibfield{author}{\bibinfo{person}{Iván~Sánchez Milara}, \bibinfo{person}{Georgi~V. Georgiev}, \bibinfo{person}{Jani Ylioja}, \bibinfo{person}{Onnur Özüduru}, {and} \bibinfo{person}{Jukka Riekki}.} \bibinfo{year}{2019}\natexlab{}.
\newblock \showarticletitle{"{Document}-while-doing": a documentation tool for {Fab} {Lab} environments}.
\newblock \bibinfo{journal}{\emph{The Design Journal}} \bibinfo{volume}{22}, \bibinfo{number}{sup1} (\bibinfo{date}{April} \bibinfo{year}{2019}), \bibinfo{pages}{2019--2030}.
\newblock
\showISSN{1460-6925, 1756-3062}
\urldef\tempurl%
\url{https://doi.org/10.1080/14606925.2019.1594926}
\showDOI{\tempurl}


\bibitem[Mildenhall et~al\mbox{.}(2020)]%
        {mildenhall_nerf_2020}
\bibfield{author}{\bibinfo{person}{Ben Mildenhall}, \bibinfo{person}{Pratul~P. Srinivasan}, \bibinfo{person}{Matthew Tancik}, \bibinfo{person}{Jonathan~T. Barron}, \bibinfo{person}{Ravi Ramamoorthi}, {and} \bibinfo{person}{Ren Ng}.} \bibinfo{year}{2020}\natexlab{}.
\newblock \bibinfo{title}{{NeRF}: {Representing} {Scenes} as {Neural} {Radiance} {Fields} for {View} {Synthesis}}.
\newblock
\newblock
\urldef\tempurl%
\url{http://arxiv.org/abs/2003.08934}
\showURL{%
\tempurl}
\newblock
\shownote{arXiv:2003.08934 [cs]}.


\bibitem[Munoz-Salinas(2012)]%
        {munoz2012aruco}
\bibfield{author}{\bibinfo{person}{Rafael Munoz-Salinas}.} \bibinfo{year}{2012}\natexlab{}.
\newblock \showarticletitle{Aruco: a minimal library for augmented reality applications based on opencv}.
\newblock \bibinfo{journal}{\emph{Universidad de C{\'o}rdoba}}  \bibinfo{volume}{386} (\bibinfo{year}{2012}).
\newblock


\bibitem[Müller et~al\mbox{.}(2022)]%
        {muller_instant_2022}
\bibfield{author}{\bibinfo{person}{Thomas Müller}, \bibinfo{person}{Alex Evans}, \bibinfo{person}{Christoph Schied}, {and} \bibinfo{person}{Alexander Keller}.} \bibinfo{year}{2022}\natexlab{}.
\newblock \showarticletitle{Instant {NGP}}.
\newblock \bibinfo{journal}{\emph{ACM Transactions on Graphics}} \bibinfo{volume}{41}, \bibinfo{number}{4} (\bibinfo{date}{July} \bibinfo{year}{2022}), \bibinfo{pages}{1--15}.
\newblock
\showISSN{0730-0301, 1557-7368}
\urldef\tempurl%
\url{https://doi.org/10.1145/3528223.3530127}
\showDOI{\tempurl}


\bibitem[Oda et~al\mbox{.}(2015)]%
        {feiner-virtualreplicas}
\bibfield{author}{\bibinfo{person}{Ohan Oda}, \bibinfo{person}{Carmine Elvezio}, \bibinfo{person}{Mengu Sukan}, \bibinfo{person}{Steven Feiner}, {and} \bibinfo{person}{Barbara Tversky}.} \bibinfo{year}{2015}\natexlab{}.
\newblock \showarticletitle{Virtual Replicas for Remote Assistance in Virtual and Augmented Reality}. In \bibinfo{booktitle}{\emph{Proceedings of the 28th Annual ACM Symposium on User Interface Software \& Technology}} (Charlotte, NC, USA) \emph{(\bibinfo{series}{UIST '15})}. \bibinfo{publisher}{Association for Computing Machinery}, \bibinfo{address}{New York, NY, USA}, \bibinfo{pages}{405–415}.
\newblock
\showISBNx{9781450337793}
\urldef\tempurl%
\url{https://doi.org/10.1145/2807442.2807497}
\showDOI{\tempurl}


\bibitem[Olson(2011)]%
        {olson_apriltag_2011}
\bibfield{author}{\bibinfo{person}{Edwin Olson}.} \bibinfo{year}{2011}\natexlab{}.
\newblock \showarticletitle{{AprilTag}: {A} robust and flexible visual fiducial system}. In \bibinfo{booktitle}{\emph{2011 {IEEE} {International} {Conference} on {Robotics} and {Automation}}}. \bibinfo{publisher}{IEEE}, \bibinfo{address}{Shanghai, China}, \bibinfo{pages}{3400--3407}.
\newblock
\showISBNx{978-1-61284-386-5}
\urldef\tempurl%
\url{https://doi.org/10.1109/ICRA.2011.5979561}
\showDOI{\tempurl}


\bibitem[ORR(1996)]%
        {orr-talking-about-machines}
\bibfield{author}{\bibinfo{person}{JULIAN~E. ORR}.} \bibinfo{year}{1996}\natexlab{}.
\newblock \bibinfo{booktitle}{\emph{Talking about Machines: An Ethnography of a Modern Job}}.
\newblock \bibinfo{publisher}{Cornell University Press}.
\newblock
\showISBNx{9780801432972}
\urldef\tempurl%
\url{http://www.jstor.org/stable/10.7591/j.ctt1hhfnkz}
\showURL{%
\tempurl}


\bibitem[Posner et~al\mbox{.}(1993)]%
        {posner1993people}
\bibfield{author}{\bibinfo{person}{Ilona~R Posner}, \bibinfo{person}{Ronald~M Baecker}, {and} \bibinfo{person}{M Mantei}.} \bibinfo{year}{1993}\natexlab{}.
\newblock \showarticletitle{How people write together}. In \bibinfo{booktitle}{\emph{Proceedings of the Hawaii International Conference on System Sciences}}, Vol.~\bibinfo{volume}{25}. IEEE INSTITUTE OF ELECTRICAL AND ELECTRONICS, \bibinfo{pages}{127--127}.
\newblock


\bibitem[Radu et~al\mbox{.}(2021)]%
        {needs-survey-AR-collab-cscw}
\bibfield{author}{\bibinfo{person}{Iulian Radu}, \bibinfo{person}{Tugce Joy}, \bibinfo{person}{Yiran Bowman}, \bibinfo{person}{Ian Bott}, {and} \bibinfo{person}{Bertrand Schneider}.} \bibinfo{year}{2021}\natexlab{}.
\newblock \showarticletitle{A Survey of Needs and Features for Augmented Reality Collaborations in Collocated Spaces}.
\newblock \bibinfo{journal}{\emph{Proc. ACM Hum.-Comput. Interact.}} \bibinfo{volume}{5}, \bibinfo{number}{CSCW1}, Article \bibinfo{articleno}{169} (\bibinfo{date}{apr} \bibinfo{year}{2021}), \bibinfo{numpages}{21}~pages.
\newblock
\urldef\tempurl%
\url{https://doi.org/10.1145/3449243}
\showDOI{\tempurl}


\bibitem[Ratcliffe(2013)]%
        {flailingmonkey2013}
\bibfield{author}{\bibinfo{person}{Mike Ratcliffe}.} \bibinfo{year}{2013}\natexlab{}.
\newblock \bibinfo{title}{The History of Firebug}.
\newblock
\newblock
\urldef\tempurl%
\url{https://flailingmonkey.com/the-history-of-firebug}
\showURL{%
\tempurl}


\bibitem[Repair(2019)]%
        {extruderJam}
\bibfield{author}{\bibinfo{person}{LA~3D~Printer Repair}.} \bibinfo{year}{2019}\natexlab{}.
\newblock \bibinfo{title}{Prusa MK3S fixing stuck filament or bad unload}.
\newblock
\newblock
\urldef\tempurl%
\url{https://www.youtube.com/watch?v=i5xnAQ5dHVs}
\showURL{%
\tempurl}


\bibitem[Sakashita et~al\mbox{.}(2023)]%
        {sakashita_vroxy_2023}
\bibfield{author}{\bibinfo{person}{Mose Sakashita}, \bibinfo{person}{Hyunju Kim}, \bibinfo{person}{Brandon Woodard}, \bibinfo{person}{Ruidong Zhang}, {and} \bibinfo{person}{François Guimbretière}.} \bibinfo{year}{2023}\natexlab{}.
\newblock \showarticletitle{{VRoxy}: {Wide}-{Area} {Collaboration} {From} an {Office} {Using} a {VR}-{Driven} {Robotic} {Proxy}}. In \bibinfo{booktitle}{\emph{Proceedings of the 36th {Annual} {ACM} {Symposium} on {User} {Interface} {Software} and {Technology}}}. \bibinfo{publisher}{ACM}, \bibinfo{address}{San Francisco CA USA}, \bibinfo{pages}{1--13}.
\newblock
\showISBNx{9798400701320}
\urldef\tempurl%
\url{https://doi.org/10.1145/3586183.3606743}
\showDOI{\tempurl}


\bibitem[Sch\"{o}nberger and Frahm(2016)]%
        {schoenberger2016sfm}
\bibfield{author}{\bibinfo{person}{Johannes~Lutz Sch\"{o}nberger} {and} \bibinfo{person}{Jan-Michael Frahm}.} \bibinfo{year}{2016}\natexlab{}.
\newblock \showarticletitle{Structure-from-Motion Revisited}. In \bibinfo{booktitle}{\emph{Conference on Computer Vision and Pattern Recognition (CVPR)}}.
\newblock


\bibitem[Sch\"{o}nberger et~al\mbox{.}(2016)]%
        {schoenberger2016mvs}
\bibfield{author}{\bibinfo{person}{Johannes~Lutz Sch\"{o}nberger}, \bibinfo{person}{Enliang Zheng}, \bibinfo{person}{Marc Pollefeys}, {and} \bibinfo{person}{Jan-Michael Frahm}.} \bibinfo{year}{2016}\natexlab{}.
\newblock \showarticletitle{Pixelwise View Selection for Unstructured Multi-View Stereo}. In \bibinfo{booktitle}{\emph{European Conference on Computer Vision (ECCV)}}.
\newblock


\bibitem[Snavely et~al\mbox{.}(2006)]%
        {snavely_photo_2006}
\bibfield{author}{\bibinfo{person}{Noah Snavely}, \bibinfo{person}{Steven~M. Seitz}, {and} \bibinfo{person}{Richard Szeliski}.} \bibinfo{year}{2006}\natexlab{}.
\newblock \showarticletitle{Photo tourism: exploring photo collections in {3D}}. In \bibinfo{booktitle}{\emph{{ACM} {SIGGRAPH} 2006 {Papers} on - {SIGGRAPH} '06}}. \bibinfo{publisher}{ACM Press}, \bibinfo{address}{Boston, Massachusetts}, \bibinfo{pages}{835}.
\newblock
\showISBNx{978-1-59593-364-5}
\urldef\tempurl%
\url{https://doi.org/10.1145/1179352.1141964}
\showDOI{\tempurl}


\bibitem[Star and Griesemer({[n.\,d.]})]%
        {star_boundary_objects}
\bibfield{author}{\bibinfo{person}{Susan~Leigh Star} {and} \bibinfo{person}{James~R. Griesemer}.} \bibinfo{year}{[n.\,d.]}\natexlab{}.
\newblock \bibinfo{title}{Institutional {Ecology}, `{Translations}' and {Boundary} {Objects}: {Amateurs} and {Professionals} in {Berkeley}'s {Museum} of {Vertebrate} {Zoology}, 1907-39}.
\newblock
\newblock
\urldef\tempurl%
\url{https://doi.org/10.1177/030631289019003001}
\showDOI{\tempurl}


\bibitem[Subbaraman and Peek(2023a)]%
        {subbaraman_3d_2023}
\bibfield{author}{\bibinfo{person}{Blair Subbaraman} {and} \bibinfo{person}{Nadya Peek}.} \bibinfo{year}{2023}\natexlab{a}.
\newblock \showarticletitle{{3D} {Printers} {Don}’t {Fix} {Themselves}: {How} {Maintenance} is {Part} of {Digital} {Fabrication}}. In \bibinfo{booktitle}{\emph{Proceedings of the 2023 {ACM} {Designing} {Interactive} {Systems} {Conference}}}. \bibinfo{publisher}{ACM}, \bibinfo{address}{Pittsburgh PA USA}, \bibinfo{pages}{2050--2065}.
\newblock
\showISBNx{978-1-4503-9893-0}
\urldef\tempurl%
\url{https://doi.org/10.1145/3563657.3595991}
\showDOI{\tempurl}


\bibitem[Subbaraman and Peek(2023b)]%
        {subbaraman-maintenance}
\bibfield{author}{\bibinfo{person}{Blair Subbaraman} {and} \bibinfo{person}{Nadya Peek}.} \bibinfo{year}{2023}\natexlab{b}.
\newblock \showarticletitle{3D Printers Don’t Fix Themselves: How Maintenance is Part of Digital Fabrication}. In \bibinfo{booktitle}{\emph{Proceedings of the 2023 ACM Designing Interactive Systems Conference}} (Pittsburgh, PA, USA) \emph{(\bibinfo{series}{DIS '23})}. \bibinfo{publisher}{Association for Computing Machinery}, \bibinfo{address}{New York, NY, USA}, \bibinfo{pages}{2050–2065}.
\newblock
\showISBNx{9781450398930}
\urldef\tempurl%
\url{https://doi.org/10.1145/3563657.3595991}
\showDOI{\tempurl}


\bibitem[Team({[n.\,d.]})]%
        {kicad}
\bibfield{author}{\bibinfo{person}{KiCad~Development Team}.} \bibinfo{year}{[n.\,d.]}\natexlab{}.
\newblock \bibinfo{title}{{KiCad} {EDA} - {Schematic} {Capture} \& {PCB} {Design} {Software}}.
\newblock
\newblock
\urldef\tempurl%
\url{https://www.kicad.org/}
\showURL{%
\tempurl}


\bibitem[Teo et~al\mbox{.}(2019)]%
        {Theophilus-360video-3Dreconstruction}
\bibfield{author}{\bibinfo{person}{Theophilus Teo}, \bibinfo{person}{Louise Lawrence}, \bibinfo{person}{Gun~A. Lee}, \bibinfo{person}{Mark Billinghurst}, {and} \bibinfo{person}{Matt Adcock}.} \bibinfo{year}{2019}\natexlab{}.
\newblock \showarticletitle{Mixed Reality Remote Collaboration Combining 360 Video and 3D Reconstruction}. In \bibinfo{booktitle}{\emph{Proceedings of the 2019 CHI Conference on Human Factors in Computing Systems}} (Glasgow, Scotland Uk) \emph{(\bibinfo{series}{CHI '19})}. \bibinfo{publisher}{Association for Computing Machinery}, \bibinfo{address}{New York, NY, USA}, \bibinfo{pages}{1–14}.
\newblock
\showISBNx{9781450359702}
\urldef\tempurl%
\url{https://doi.org/10.1145/3290605.3300431}
\showDOI{\tempurl}


\bibitem[Thompson(2020)]%
        {thompson2020coders}
\bibfield{author}{\bibinfo{person}{Clive Thompson}.} \bibinfo{year}{2020}\natexlab{}.
\newblock \bibinfo{booktitle}{\emph{Coders: The making of a new tribe and the remaking of the world}}.
\newblock \bibinfo{publisher}{Penguin}.
\newblock


\bibitem[Thoravi~Kumaravel et~al\mbox{.}(2020)]%
        {transceiVR-VR+externals}
\bibfield{author}{\bibinfo{person}{Balasaravanan Thoravi~Kumaravel}, \bibinfo{person}{Cuong Nguyen}, \bibinfo{person}{Stephen DiVerdi}, {and} \bibinfo{person}{Bjoern Hartmann}.} \bibinfo{year}{2020}\natexlab{}.
\newblock \showarticletitle{TransceiVR: Bridging Asymmetrical Communication Between VR Users and External Collaborators}. In \bibinfo{booktitle}{\emph{Proceedings of the 33rd Annual ACM Symposium on User Interface Software and Technology}} (Virtual Event, USA) \emph{(\bibinfo{series}{UIST '20})}. \bibinfo{publisher}{Association for Computing Machinery}, \bibinfo{address}{New York, NY, USA}, \bibinfo{pages}{182–195}.
\newblock
\showISBNx{9781450375146}
\urldef\tempurl%
\url{https://doi.org/10.1145/3379337.3415827}
\showDOI{\tempurl}


\bibitem[Tran~O'Leary et~al\mbox{.}(2024)]%
        {tran_oleary_tandem_2024}
\bibfield{author}{\bibinfo{person}{Jasper Tran~O'Leary}, \bibinfo{person}{Thrisha Ramesh}, \bibinfo{person}{Octi Zhang}, {and} \bibinfo{person}{Nadya Peek}.} \bibinfo{year}{2024}\natexlab{}.
\newblock \showarticletitle{Tandem: {Reproducible} {Digital} {Fabrication} {Workflows} as {Multimodal} {Programs}}. In \bibinfo{booktitle}{\emph{Proceedings of the {CHI} {Conference} on {Human} {Factors} in {Computing} {Systems}}}. \bibinfo{publisher}{ACM}, \bibinfo{address}{Honolulu HI USA}, \bibinfo{pages}{1--16}.
\newblock
\showISBNx{9798400703300}
\urldef\tempurl%
\url{https://doi.org/10.1145/3613904.3642751}
\showDOI{\tempurl}


\bibitem[Utz and Wolfers(2022)]%
        {utz2022videos}
\bibfield{author}{\bibinfo{person}{Sonja Utz} {and} \bibinfo{person}{Lara~N Wolfers}.} \bibinfo{year}{2022}\natexlab{}.
\newblock \showarticletitle{How-to videos on YouTube: The role of the instructor}.
\newblock \bibinfo{journal}{\emph{Information, Communication \& Society}} \bibinfo{volume}{25}, \bibinfo{number}{7} (\bibinfo{year}{2022}), \bibinfo{pages}{959--974}.
\newblock


\bibitem[Wagstaff(2012)]%
        {Wagstaff_2012}
\bibfield{author}{\bibinfo{person}{Keith Wagstaff}.} \bibinfo{year}{2012}\natexlab{}.
\newblock \bibinfo{title}{IKEA Starts ‘How to Build’ YouTube Channel to Help Frustrated Customers | TIME.com}.
\newblock
\newblock
\urldef\tempurl%
\url{https://techland.time.com/2012/02/24/ikea-starts-how-to-build-youtube-channel-to-help-frustrated-customers/}
\showURL{%
\tempurl}


\bibitem[Wang et~al\mbox{.}(2021)]%
        {wang2021ar}
\bibfield{author}{\bibinfo{person}{Peng Wang}, \bibinfo{person}{Xiaoliang Bai}, \bibinfo{person}{Mark Billinghurst}, \bibinfo{person}{Shusheng Zhang}, \bibinfo{person}{Xiangyu Zhang}, \bibinfo{person}{Shuxia Wang}, \bibinfo{person}{Weiping He}, \bibinfo{person}{Yuxiang Yan}, {and} \bibinfo{person}{Hongyu Ji}.} \bibinfo{year}{2021}\natexlab{}.
\newblock \showarticletitle{AR/MR remote collaboration on physical tasks: A review}.
\newblock \bibinfo{journal}{\emph{Robotics and Computer-Integrated Manufacturing}}  \bibinfo{volume}{72} (\bibinfo{year}{2021}), \bibinfo{pages}{102071}.
\newblock


\bibitem[Wang et~al\mbox{.}({[n.\,d.]})]%
        {wang_dust3r_nodate}
\bibfield{author}{\bibinfo{person}{Shuzhe Wang}, \bibinfo{person}{Vincent Leroy}, \bibinfo{person}{Yohann Cabon}, \bibinfo{person}{Boris Chidlovskii}, {and} \bibinfo{person}{Jerome Revaud}.} \bibinfo{year}{[n.\,d.]}\natexlab{}.
\newblock \showarticletitle{{DUSt3R}: {Geometric} {3D} {Vision} {Made} {Easy}}.
\newblock  (\bibinfo{year}{[n.\,d.]}).
\newblock


\bibitem[Yamauchi et~al\mbox{.}(2000)]%
        {yamauchi_collaboration_2000}
\bibfield{author}{\bibinfo{person}{Yutaka Yamauchi}, \bibinfo{person}{Makoto Yokozawa}, \bibinfo{person}{Takeshi Shinohara}, {and} \bibinfo{person}{Toru Ishida}.} \bibinfo{year}{2000}\natexlab{}.
\newblock \showarticletitle{Collaboration with {Lean} {Media}: how open-source software succeeds}. In \bibinfo{booktitle}{\emph{Proceedings of the 2000 {ACM} conference on {Computer} supported cooperative work}}. \bibinfo{publisher}{ACM}, \bibinfo{address}{Philadelphia Pennsylvania USA}, \bibinfo{pages}{329--338}.
\newblock
\showISBNx{978-1-58113-222-9}
\urldef\tempurl%
\url{https://doi.org/10.1145/358916.359004}
\showDOI{\tempurl}


\bibitem[Yevstifeyev(2011)]%
        {viewsource}
\bibfield{author}{\bibinfo{person}{Mykyta Yevstifeyev}.} \bibinfo{year}{2011}\natexlab{}.
\newblock \bibinfo{booktitle}{\emph{{The 'view-source' URI Scheme}}}.
\newblock \bibinfo{type}{Internet-Draft} draft-yevstifeyev-view-source-uri-01. \bibinfo{institution}{Internet Engineering Task Force}.
\newblock
\urldef\tempurl%
\url{https://datatracker.ietf.org/doc/draft-yevstifeyev-view-source-uri/01/}
\showURL{%
\tempurl}
\newblock
\shownote{Work in Progress}.


\bibitem[Yu et~al\mbox{.}(2022)]%
        {yu2022duplicated}
\bibfield{author}{\bibinfo{person}{Kevin Yu}, \bibinfo{person}{Ulrich Eck}, \bibinfo{person}{Frieder Pankratz}, \bibinfo{person}{Marc Lazarovici}, \bibinfo{person}{Dirk Wilhelm}, {and} \bibinfo{person}{Nassir Navab}.} \bibinfo{year}{2022}\natexlab{}.
\newblock \showarticletitle{Duplicated reality for co-located augmented reality collaboration}.
\newblock \bibinfo{journal}{\emph{IEEE Transactions on Visualization and Computer Graphics}} \bibinfo{volume}{28}, \bibinfo{number}{5} (\bibinfo{year}{2022}), \bibinfo{pages}{2190--2200}.
\newblock


\end{thebibliography}

\end{document}